\pdfoutput=1
\documentclass[sigplan,10pt,review=false,nonacm=true]{acmart}
\usepackage{graphicx}
\usepackage{subcaption}
\usepackage{algorithm}
\usepackage[noend]{algpseudocode}
\usepackage{mathtools}
\usepackage{tikz}
\usepackage{mleftright}
\usepackage{nicefrac}
\usepackage{varwidth}
\usepackage{comment}
\usepackage{multirow}
\usepackage{gnuplottex}
\renewcommand\footnotetextcopyrightpermission[1]{} 

\algnewcommand{\LineComment}[1]{\State \(\triangleright\) #1}
\algnewcommand\And{\textbf{and}}
\renewcommand*\Call[2]{\textproc{#1}(#2)}
\algnewcommand\algorithmicforeach{\textbf{for each}}
\algdef{S}[FOR]{ForEach}[1]{\algorithmicforeach\ #1\ \algorithmicdo}

\newcommand{\rot}[1]{\rotatebox{90}{#1}}

\AtBeginDocument{%
  \providecommand\BibTeX{{%
    \normalfont B\kern-0.5em{\scshape i\kern-0.25em b}\kern-0.8em\TeX}}}

\setcopyright{acmcopyright}
\copyrightyear{2021}
\acmYear{2021}

\newcommand{\ralg}[1]{Algorithm~\ref{alg:#1}}
\newcommand{\rsec}[1]{Section~\ref{sec:#1}}

\newcommand{\rtab}[1]{Table~\ref{tab:#1}}
\newcommand{\rfig}[1]{Figure~\ref{fig:#1}}
\newcommand{\rfigs}[2]{Figures~\ref{fig:#1} --~\ref{fig:#2}}
\newcommand{\rlst}[1]{Listing~\ref{lst:#1}}

\newcommand*\circled[1]{\tikz[baseline=(char.base)]{
\node[shape=circle,draw,inner sep=1pt] (char) {#1};}}

\usepackage{listings}
\usepackage{xcolor}

\definecolor{codepurple}{rgb}{0.58,0,0.82}
\definecolor{backcolour}{rgb}{0.95,0.95,0.92}
\definecolor{darkblue}{rgb}{0.0,0.0,0.6}

\definecolor{mGreen}{rgb}{0,0.6,0}
\definecolor{mGray}{rgb}{0.5,0.5,0.5}
\definecolor{mPurple}{rgb}{0.58,0,0.82}
\definecolor{backgroundColour}{rgb}{0.95,0.95,0.92}
\definecolor{backgroundColour}{rgb}{0.95,0.95,0.92}

\lstdefinestyle{mystyle}{
  commentstyle=\foonotesize\color{mGreen},
  backgroundcolor=\color{backcolour},
  stringstyle=\color{mPurple},
  basicstyle=\footnotesize,
  breakatwhitespace=false,
  breaklines=true,
  captionpos=b,
  keepspaces=true,
  showspaces=false,
  showstringspaces=false,
  showtabs=false,
  tabsize=2,
  numberstyle=\tiny\color{mGray},
  numbers=left,                    
  numbersep=5pt,    
  float=tp,
  floatplacement=tbp
}

\lstdefinelanguage{mlir}{
  alsoletter={\%!^.\<\>_},
  alsoother={\(\)\,:},
  morecomment=[l]{//}, 
  morestring=[b]", 
  sensitive=true,
  keywords={\%arg0,\%arg1,\%c1_i32,\%c0_i32,\%0,\%1,\%2,\%3,\%4,\%5},
  keywords=[2]{i32, i64},
  moredelim=[s][keywordstyle2]{!}{\>},
}

\lstdefinestyle{mlir}{
  commentstyle=\foonotesize\color{mPurple},
  backgroundcolor=\color{backcolour},
  stringstyle=\color{darkblue},
  keywordstyle=[1]{\color{orange}},
  keywordstyle=[2]{\color{red}},
  basicstyle=\footnotesize,
  breakatwhitespace=false,
  breaklines=true,
  captionpos=b,
  keepspaces=true,
  showspaces=false,
  showstringspaces=false,
  showtabs=false,
  tabsize=2,
  numberstyle=\tiny\color{mGray},
  numbers=left,                    
  numbersep=5pt,    
  float=tp,
  floatplacement=tbp,
  rulecolor=\color{black}
}

\hypersetup{%
  colorlinks=true,
  linkcolor=blue,
  linkbordercolor=blue,
}

\begin{document}

\title{Source Matching and Rewriting}

\author{
Vinicius Couto, Luciano Zago, Hervé Yviquel and Guido Araujo
}
\affiliation{\institution{Institute of Computing - UNICAMP \city{Campinas} \country{Brazil}}}
\email{{v188115,l182835}@dac.unicamp.br}
\email{{hyviquel,guido}@unicamp.br}

\begin{abstract}
A typical compiler flow relies on a uni-directional sequence of translation/optimization steps that \textit{lower} the program abstract representation, making it hard to preserve higher-level program information across each transformation step. On the other hand, modern ISA extensions and hardware accelerators can benefit from the compiler’s ability to detect and \textit{raise} program idioms to acceleration instructions or optimized library calls. Although recent works based on Multi-Level IR (MLIR) have been proposed for code raising, they rely on specialized languages, compiler recompilation, or in-depth dialect knowledge. This paper presents \textit{Source Matching and Rewriting} (SMR), a user-oriented source-code-based approach for MLIR idiom matching and rewriting that does not require a compiler expert’s intervention. SMR uses a two-phase automaton-based DAG-matching algorithm inspired by early work on tree-pattern matching. First, the idiom \textit{Control-Dependency Graph} (CDG) is matched against the program’s CDG to rule out code fragments that do not have a control-flow structure similar to the desired idiom. Second, candidate code fragments from the previous phase have their \textit{Data-Dependency Graphs} (DDGs) constructed and matched against the idiom DDG. Experimental results show that SMR can effectively match idioms from Fortran (FIR) and C (CIL) programs while raising them as BLAS calls to improve performance.
\end{abstract}

\maketitle

\section{Introduction}
\label{sec:introduction}
Idiom recognition is a well-known and studied problem in computer science, which aims to identify program fragments~\cite{Blume1995, Pottenger1995, Lee2004, Arenaz2008, Ginsbach2018, Ginsbach2020}.  Although idiom recognition has found a niche in compiling technology, in areas like code generation (e.g., instruction selection)~\cite{Aho1976,Pelegri-Llopart1988,Aho1989,Fraser1992}, its broad application is considerably constrained by how modern compilers work. A typical compiler flow makes a series of translation passes that lowers the level of abstraction from source to machine code, with the goal of  optimizing the code at each level. One such example is the Clang/LLVM compiler~\cite{Lattner} which starts at a \textit{high-level} language (e.g., C or Fortran) and gradually lowers the abstraction level to AST, LLVM IR, Machine IR, and finally  binary code. After lowering from one level to another, program information is lost, thus restricting the possibility of simultaneously combining optimization passes from different abstraction levels.  

Until recently, there was no demand for combining optimization passes from distinct abstraction levels, mainly due to two reasons. First, although many languages  claim to implement \textit{higher-level} representations, their abstraction levels are not that far above those found in  machine code (C code is an example)~\cite{Chelini2021}. Second, general-purpose processors implement very low-level computing primitives. However, the recent adoption of hardware accelerators is changing this scenario and pushing the need for optimizations and code generation at higher levels of abstractions (this also explains the renewed interest in Domain-Specific Languages). 

To use such accelerators, a compiler should lower some parts of the program while raising others.  For example, if an instruction in some ISA extension can perform  a tiled multiplication (e.g., Intel AMX or IBM MMA), the compiler should be able to raise the code to this instruction (higher-level) representation while simultaneously lowering the remaining parts of the code. A similar situation also happens when generating code for Machine Learning (ML) engines. Thus, any robust approach for idiom recognition should be able to: (a) work on a representation that enables the co-existence of different levels of  abstraction; (b) allow both raising \underline{and} lowering of the program between different levels.

\begin{lstlisting}[style=mystyle, label={lst:example}, caption={Raising dot-product and replacing by a BLAS call using SMR. Idiom and replacement are encapsulated in wrapper functions.\vspace{-0.4cm}
}]
C {
 void dot(int N,float *x,int ix,float *y,int iy,float out) {
  for (int i = 0; i  <  N;  i++) 
   out += x[i*ix] * y[i*iy];
 }
} = {
 #include <cblas.h>
 void dot(int M,float *p,int ip,float *q,int iq,float out) {
  out += cblas_sdot(M, p, ip, q, iq);
 }
}
\end{lstlisting}

Fortunately, recent research has  started to address these two requirements. To deal with (a), MLIR~\cite{Lattner2021} enables the interplay of different languages through a common Multi-Level IR. To address (b), Chelini et al.~\cite{Chelini2021} and Lücke et al.~\cite{Lucke2021} proposed techniques that raise the level of abstraction. 

This paper proposes a source code based approach for program matching and rewriting that leverages MLIR for lowering and raising. To achieve that, it  makes two assumptions. First, it assumes the existence of MLIR implementations for both matching and replacement codes. Second, it considers it impossible to automatically capture more program information beyond the one available at the source code, as subsequent  transformations reduce the  information available for analysis and optimization. Based on that, this paper claims that source code is the best abstraction level for idiom specification and a  better choice to allow the non-expert programmer to design and explore his/her own idioms.

Given the discussion above, this paper describes \textit{Source Matching and Re-writing} (SMR), a system that relies on a small declarative language (PAT) to match and re-write program code fragments using MLIR  as a supporting framework. In SMR, the \textit{input} and the \textit{idiom} to be identified  are lowered to their  corresponding MLIR dialects (e.g., FIR~\cite{fir} or CIL~\cite{cil}). Then, an approach inspired by early work on tree-rewriting systems~\cite{Aho1989} is used to match  the MLIR idiom pattern against the input MLIR to enable code lowering or raising. 

Consider, for example, the PAT description in \rlst{example} that is divided into two code sections. The first section (lines 1--6) describes the idiom code. The letter "C" at the beginning of the first section (line 1) instructs SMR that the idiom to be detected was written in the C language. In a PAT description, the idiom is encapsulated as a function (lines 2--5) where the arguments are the inputs of the idiom (in this case, dot-product). The second section (lines 6--11) describes the replacement code. It is also declared as a function (lines 8--10) with the same signature as the idiom  that it intends to replace (i.e., dot-product). As detailed in the following sections, a PAT description could potentially use any language that can be compiled to MLIR and integrated with SMR.  Nevertheless, inter-language/dialect rewrites are not explored in this paper. As future work, we intend to research how to leverage MLIR inter-dialect functionalities to abstract inter-language rewrites in SMR.

This  paper is divided as follows. \rsec{background} provides a background of the techniques used in the SMR design. \rsec{smr} gives an overview of the SMR architecture, and \rsec{smralgo} details its main algorithms. \rsec{related_works} reviews other works related to this paper. \rsec{results} shows the experimental results, and \rsec{conclusions} concludes the work.

\vspace{-0.2cm}
\section{Background} 
\label{sec:background}

One of the central tasks in idiom recognition is the  ability to pattern match the input program. Pattern matching has been extensively used to design instruction selection algorithms needed in the code generation phase of a compiler~\cite{Aho1976,Pelegri-Llopart1988,Aho1989,Fraser1992}.   

The work proposed herein for idiom recognition (SMR) uses a similar approach as instruction selection but differs in two major aspects.  First, instead of representing patterns and the input program in a low-level IR, SMR uses MLIR. This was motivated by the fact that MLIR enables a common IR structure~\cite{mlir-docs} between distinct source languages.  For example, in this paper, both Fortran and C source codes were lowered to their respective MLIR dialects (FIR and CIL), from which matching and re-writing can be performed. Second, contrary to instruction selection, SMR uses DAG and not tree matching (\rsec{smr}) and applies it for both control and data flow matching. 

To better understand the algorithms proposed by SMR, this section covers background material on MLIR (\rsec{mlir}) and tree-pattern matching (\rsec{treematching}). 

\subsection{The Multi-Level Intermediate Representation}
\label{sec:mlir}

MLIR~\cite{Lattner2021} is a \textit{framework} with several tools for building complex and reusable compilers. Its key idea is a hybrid intermediate representation that can support different levels of abstraction. MLIR uses an interface that relies on a set of basic declarative elements~\cite{mlir-docs}: \textit{Operations, Attributes, Regions, Basic Blocks}, and re-writing rules. Such elements can be configured to form a \textit{dialect} with a specific abstraction-level representation.

An MLIR dialect can be generated from source code with the goal of preserving high-level language information during the compilation flow. For example, C and Fortran can use MLIR to respectively translate Clang and Flang AST to the FIR~\cite{fir} and CIL~\cite{cil} dialects.

Although each dialect is somewhat unique, all of them must follow a common set of rules imposed by the MLIR interface~\cite{mlir-docs}. The different syntax concepts of the language are implemented by configuring this interface. For example, source code \texttt{if-else} clauses from distinct languages can be implemented by configuring MLIR concepts of \textit{regions} and \textit{operations}. As discussed in \rsec{smr}, SMR relies on the common rules/structures imposed by MLIR's interface~\cite{mlir-docs} (while also allowing some configuration of its own on a dialect basis) to design a single algorithm that performs idiom matching and rewriting for both CIL and FIR inputs.

To define their structures and rewrite patterns, MLIR \textit{Operations} are modeled using the Table-Gen-based~\cite{LLVMDocumentation2021} specification for Operations Descriptions (ODS) that is extensively used in LLVM. MLIR ODS description is eventually translated to C++ code which is then integrated into the rest of the system. Some of the basic declarative elements of MLIR are described below.

\begin{itemize}
\item \textbf{Operations}\\
Operations in MLIR are similar to those in other IRs instructions: an operation may have input operands, generating an output operand. However, it can also contain attributes, a list of regions, and multiple output operands, among other elements.
     
\item \textbf{Attributes}\\
Attributes are used to identify specific features of operations. An operation that initializes a constant, for example, may have an attribute that defines its value, as is the case with the "\texttt{std.const}" operation, which uses the "\texttt{value}" attribute for this purpose.

\item \textbf{Basic Blocks}\\
\textit{MLIR Basic Blocks} (MLIR BBs) are defined similarly as in other IRs with a few differences. First, contrary to the classical definition in \cite{Lam2007,Cooper2012}, MLIR BBs have operations that can contain MLIR regions. Another difference is the concept of \textit{basic block arguments}. These arguments abstract control-flow dependent values indicating which SSA values are available in a block. 
\item \textbf{Regions}\\
MLIR regions are lists of basic blocks that may correspond to a classical \textit{reducible} CFG region~\cite{Lam2007,Cooper2012}, or not, depending on how the language generates its corresponding MLIR representation. For example, some languages (e.g., Fortran) implement  classical reducible CFG regions (e.g., canonical loops) as an MLIR region with a single basic block, while \textit{irreducible} parts of the CFG are implemented using a region with multiple basic blocks.
\end{itemize}

\begin{figure}[!t]
    \centering
    \includegraphics[width=0.45\textwidth]
    {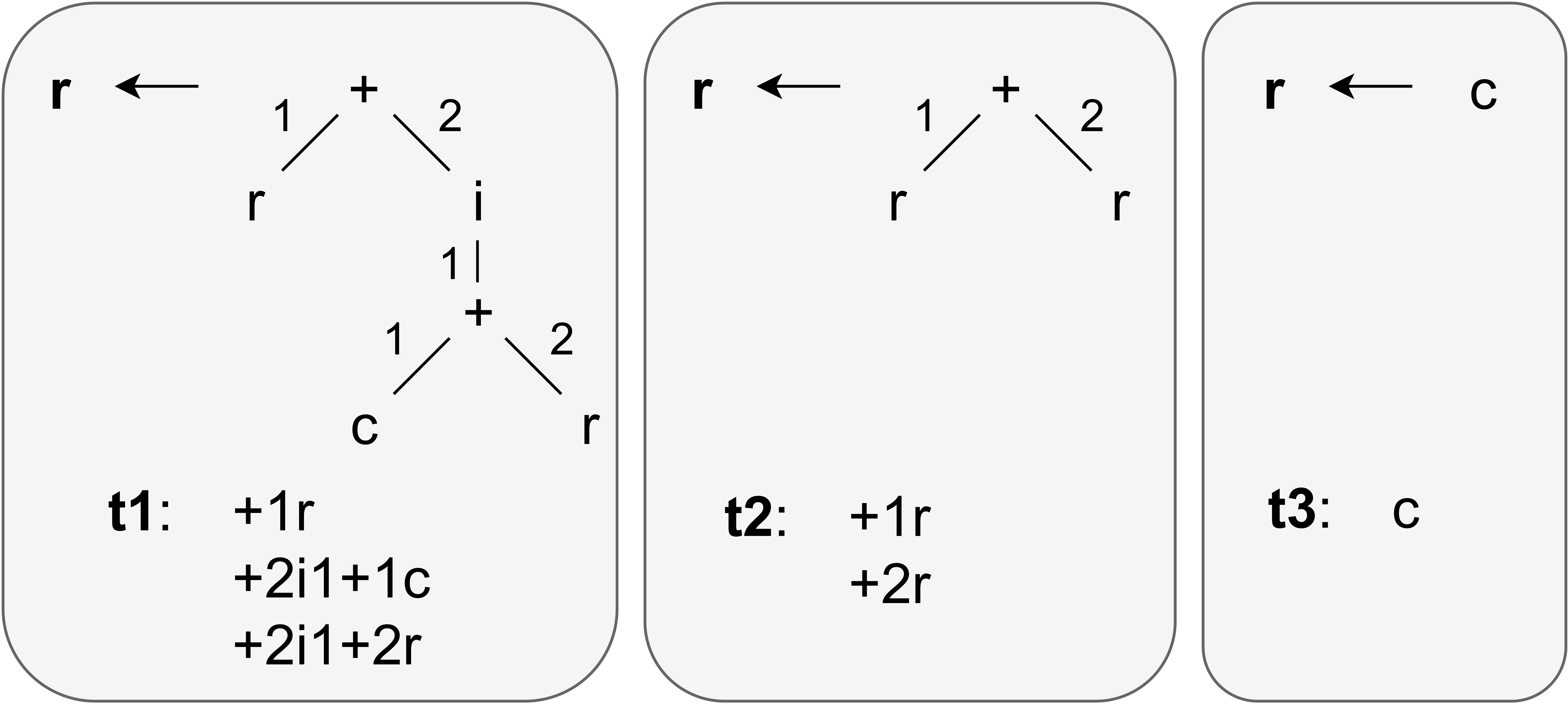}
    \caption{Tree-patterns (t1, t2 and t3). Adapted from~\cite{Aho1989}.}
    \label{fig:twigpat}
\end{figure}

\subsection{Automaton-based Tree-Matching}
\label{sec:treematching}

Instruction selection is a well-studied area in compiling technology~\cite{Lam2007}. During instruction selection, expression trees from a low-level IR of the program (e.g., LLVM Machine IR)  are covered using a library of \textit{tree patterns}. Aho et al.~\cite{Aho1989} proposed a thorough solution based on a  \textit{tree-pattern matching} algorithm that leverages Hoffman and O'Donnel~\cite{Hoffmann1982} automaton string matching approach to reduce the size of the encoded tree-patterns and improve performance. Their solution was demonstrated  in a tool called \texttt{Twig}. This paper extends~\cite{Aho1989} to enable CDG and DDG pattern matching algorithms to be used for idiom recognition in MLIR.

\begin{figure}[!t]
    \centering
    \includegraphics[width=0.45\textwidth]
    {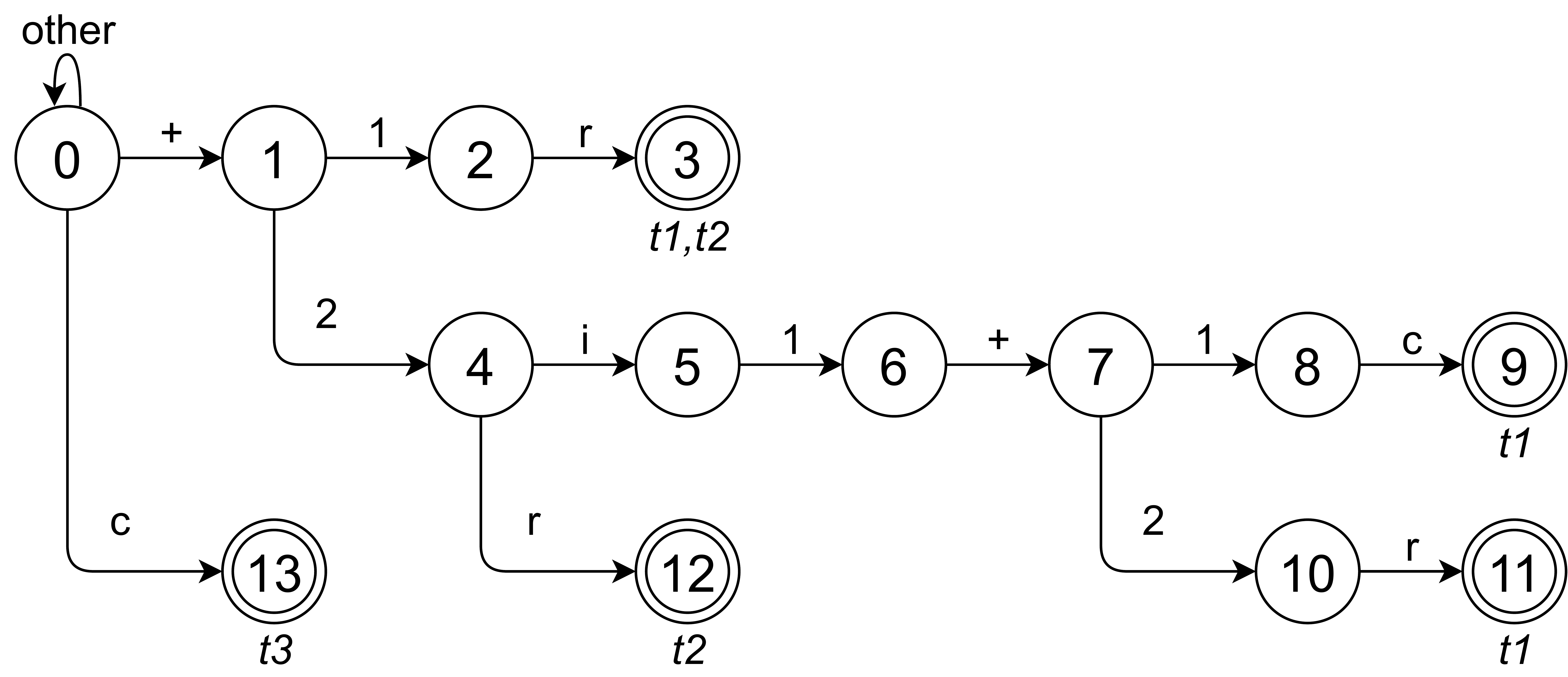}
    \caption{Tree-pattern automaton. Adapted from~\cite{Aho1989}.}
    \label{fig:twigautomaton}
\end{figure}

\rfigs{twigpat}{twigautomaton} show a quick overview on how \texttt{Twig}'s  algorithms work. First, tree-patterns \rfig{twigpat}(\texttt{t1})-(\texttt{t3}) are linearized by listing all paths from the root to a leaf. The result is a set of \textit{path-strings} encoding each tree-pattern. These strings are then converted into an Automaton (\rfig{twigautomaton}) where the transitions from states are annotated with the path-string elements, and the last state associated to the end of a path-string is marked as final. Given that strings from  different tree-patterns have common prefixes, they will traverse the same set of states in the automaton, thus reducing the automaton size. For example, consider the path-string (\texttt{+1r}) for patterns \texttt{t1} and \texttt{t2} of \rfig{twigpat}. In \rfig{twigautomaton} they share the same sequence of states \texttt{(0,1,2,3)} finishing at state \texttt{3} which is marked final (double circled), and annotated as having recognized path string \texttt{+1r} from both tree-patterns \texttt{t1} and \texttt{t2}. A tree-pattern \textit{matches} if all its path-strings end in a final state.

\vspace{-0.2cm}
\section{An Overview of SMR and PAT}
\label{sec:smr}

This section provides an overview of the SMR approach. It first describes PAT, a language that enables the user (or compiler developer) to specify an idiom and its corresponding replacement code. Second, it shows how PAT and SMR are integrated into a Clang/LLVM compilation flow. The reader should be attentive to the following notation, which will be used from now on: (a)  \textit{pattern}  is the idiom code to be matched;  (b) \textit{input} is the program inside of which SMR will detect the idiom; and (c)  \textit{replacement} is the code that will replace the matched idiom within the re-written input.

As shown in \rlst{pat}, and anticipated in \rsec{introduction}, a PAT\footnote{The EBNF representation of PAT is trivial, and thus we opted to present it in a descriptive format.} description is divided into two sections. In the first section (lines 1--5 of \rlst{pat}),  the language of the idiom to be matched (\texttt{langA}) is specified (line 1), followed by the declaration of the idiom's  \textit{wrapper function} (also written in \texttt{langA}) that declares the arguments of the idiom pattern and their corresponding types (line 2). The pattern itself is described in the body (\texttt{matchA}) of  the wrapper function (line 3). The second section of a PAT description  (lines 5--9) contains the replacement code that will rewrite the idiom when it matches some  input code fragment. Similarly, as in the idiom section, the replacement code language (\texttt{langB}) is also specified (line 5). The  wrapper function declaration (line 6, written in \texttt{langB})  matches the arguments of the idiom to the variables used by the  matched input  fragment. In the case of an idiom match, the matched input fragment is replaced by the code in \texttt{rewriteB}.

The wrapper functions in the pattern and replacement sections of a PAT description serve two, and only two, purposes. First, they make the codes valid, as the front-end must be able to compile them. Second, they  act as an interface between the input, pattern, and replacement codes. That said, the wrapper function is not a part of the pattern and will never be matched, only its arguments declarations are relevant. In the matching process, SMR considers the wrapper function arguments' order and types, as well as the function body, ignoring function's and arguments' names.

\begin{lstlisting}[floatplacement=t,style=mystyle, label={lst:pat}, caption={Replacing matchA in langA with rewriteB in langB (optional if langB = langA).}].
langA { 
  fun (type1 arg1, type2 arg2,...) { 
    matchA 
  }
} = langB {
  fun (type1 arg1, type2 arg2,...) {
    rewriteB 
  }
}
\end{lstlisting}

It is expected from the PAT file that each idiom pattern code is semantically equivalent to its respective replacement code. The author of the PAT file must ensure the correctness of such equivalence as SMR does not verify it. 

PAT differs from other proposed matching languages in many ways. One remarkable differences should be highlighted, though. In contrast to approaches such as RISE~\cite{Lucke2021}, that use their own concepts and languages, PAT describes idiom and replacement codes using regular programming languages, thus  considerably simplifying the description. 

\rfig{smr} shows how the PAT language and the SMR approach are combined with the FIR front-end. The flow for CIL's front-end is similar and uses the same SMR algorithm. Initially, two input files are provided to the compilation flow: (a) the input code that might  be rewritten~\circled{1}; and (b) a PAT file that has a list of idiom/replacement pairs~\circled{2}.

As shown in \rfig{smr}, before compiled, the PAT file is parsed~\circled{2}, separating each pattern/replacement codes into  a pair of Fortran files: \texttt{idiom.f90}  and \texttt{replacement.f90}~\circled{3}. These files, together with \texttt{input.f90} \circled{1}, are then compiled using FIR's front-end to generate  their corresponding FIR codes \circled{4}.

\vspace{-0.2cm}
\section{The SMR Algorithm}
\label{sec:smralgo}
\begin{figure}[!t]
    \centering
    \includegraphics[width=0.45\textwidth,keepaspectratio]
    {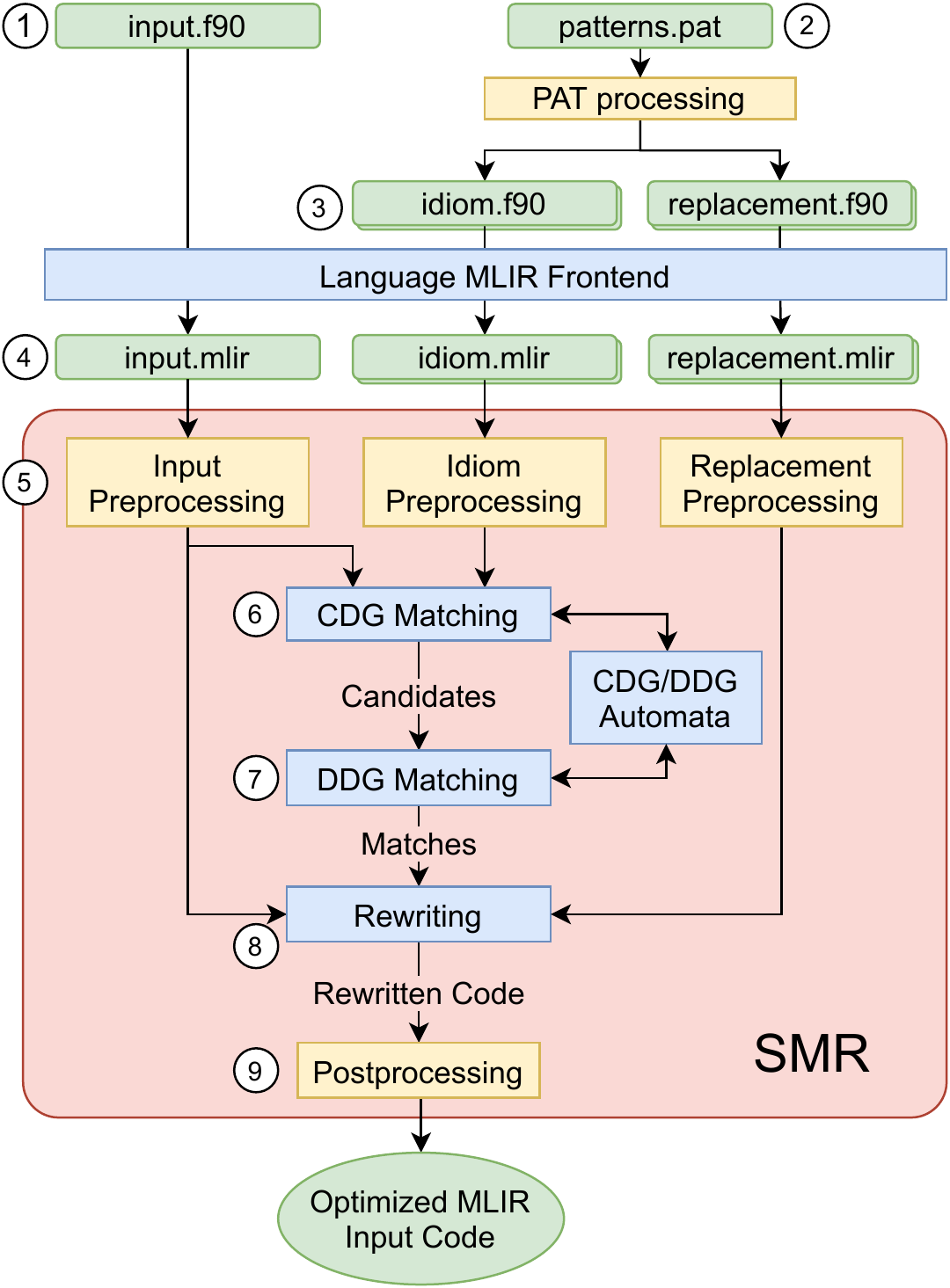}
    \caption{SMR Compilation Flow.}
    \label{fig:smr}
\end{figure}

As mentioned before, SMR relies on automaton-based DAG isomorphism to match idioms against an input program. Two major issues need to be addressed to enable that.  First, although idioms are fairly small code fragments and matching their DAGs is fast, input programs can contain thousands to millions of lines of code, making it unfeasible to pattern match idioms against a whole program. To address this, SMR uses a two-phase approach that narrows down the matching search space by (a) selecting a set of candidate fragments in the input program that has a control-flow structure similar to the one in a given idiom (\rsec{cdg}); and (b) from the filtered set of candidates, identifying those which have the same data-dependencies as the desired idiom (\rsec{ddg}). Moreover, although DAG matching is a GI-Complete~\cite{dagi} problem, idiom DDGs tend to be small and quite similar to trees, thus avoiding potential combinatorial explosions in SMR execution time. Second, depending on the problem, the number of similar idioms that can be matched could be very large, thus increasing the DAG matching algorithm's execution time and memory usage. To optimize this task, SMR encodes idiom patterns as strings and uses automatons to compress them, similarly as proposed in Aho et al.~\cite{Aho1989}. Automatons can also be easily serialized and saved for reuse, a feature that will be available in future  SMR versions. To deal with these tasks, SMR follows a compilation pipeline that implements the following sequence of operations (shown in \rfig{smr}).

\textbf{Pre/Post-processing:} \circled{5}~\circled{9} SMR offers a language-wise pre-processing pipeline to ease integration of new front-ends and dialects. It starts by receiving the MLIR code of the input program, the idiom patterns, and their corresponding replacements. The input and replacements may be optionally pre-processed in case they do not conform to the SMR constraints (\rsec{smr-limitations}). We call these tasks \textit{normalizations}, as they aim to modify the code into an SMR compliant structure without altering its computation. For now, we use normalization only to peel off the idiom code, as only its body is of interest during matching. In the future, a language-wise pre/post-processing pipeline can also be defined to deal with other requirements that might arise with the addition of new compiler front-ends.

\textbf{CDG Matching} \circled{6} After pre-processing, the idiom's MLIR code enters the \texttt{CDG Matching} pass to extract their CDG patterns and encode them as a set of strings. These sets are then used to build the CDG Automaton for matching. Next, SMR traverses the input MLIR code and encodes the input CDG as a set of strings. This set is then fed to the \texttt{CDG Automaton} to search for a control-flow match. Finally, \texttt{CDG Matching} outputs all candidates (input code fragments) that have a CDG identical to some idiom's pattern CDG. 

\textbf{DDG Matching} \circled{7} At this step, the  candidate idioms resulting from \texttt{CDG Matching} are read by the \texttt{DDG Matching} pass, which builds a set of input strings for each candidate based on  their MLIR representations.  The \texttt{DDG Matching} traverses the ud-chains of the PAT file idioms, encoding them as strings to build the \texttt{DDG Automaton}. Finally, it feeds the set of strings generated from the input candidates to the \texttt{DDG Automaton} for matching.

\textbf{Rewriting} \circled{8} After some idiom matches an input program fragment, its corresponding  MLIR code is substituted by a function call to the replacement code associated with the matched idiom pattern. A definition of such function is inserted into the input program to ensure that it is a callable function.\footnote{Function in-lining will eventually be used here.}

\rlst{idiom-f90} shows an example of a Fortran  idiom, and \rlst{idiom-cdg}, its corresponding MLIR representation. This idiom will be used in the following sections to better explain SMR, focusing particularly on the CDG/DDG Matching algorithms, as they are the two key tasks.

\begin{algorithm}[!t]
\footnotesize
\begin{algorithmic}[1]
\caption{Control-Dependency Graph Matching}
\Function{CDGMatch}{Input,Patterns}
    \LineComment  Input: MLIR input code
    \LineComment  Patterns: set of MLIR pattern codes to match
\\
    \LineComment Build automaton FSM from pattern's CDG strings
    \State $patStrings \gets []$
    \ForEach{$pat \in Patterns$}
        \LineComment pat: a single pattern code
        \State $string \gets \Call{StringyPatCDG}{pat}$
        \State $patStrings.append(string)$
    \EndFor
    \State $\Call{CDGAutomaton.build}{patStrings}$
\\
    \LineComment Generate the Input CDG string representation
    \State $inStrings \gets \Call{StringyInCDG}{Input}$
\\
    \LineComment Find input code match candidates
    \State $Candidates \gets \{\}$
    \ForEach{$inString \in inStrings$}
        \State $patternIndexes \gets \Call{CDGAutomaton.run}{inString}$
        \State $rdo \gets getRootRDO(inString)$
        \If{$patternIndexes \neq []$}
            \State $Candidates.insert(rdo)$
        \EndIf
    \EndFor
\\
    \LineComment Return RDO set with match candidates
    \State \Return Candidates
\EndFunction
\label{alg:cdg}
\end{algorithmic}
\end{algorithm}
\begin{algorithm}[!h]
\footnotesize
\begin{algorithmic}[1]
\caption{Data-Dependency Graph Matching}
\Function{DDGMatch}{Candidates, Patterns}
    \LineComment  Candidates: MLIR operations filtered by CDG matching
    \LineComment  Patterns: List of MLIR code patterns to match
\\
    \LineComment Generate the DDG stringset representation of each pattern
    \State $patStrings \gets []$
    \For{$pat \in Patterns$}
        \LineComment pat: single pattern code
        \State $ddg \gets \Call{BuildPatDDG}{pat}$
        \State $dataStringSet \gets \Call{StringfyPatDDG}{ddg}$
        \State $patStringSets.append(dataStringSet)$
    \EndFor
\\
    \LineComment Build the DDG automaton to resolve stringset matching
    \State $\Call{DDGAutomaton.build}{patStringsets}$
\\
    
    \LineComment Get which input candidates matches which patterns
    \State $matches \gets []$
    \ForEach{$rdo \in Candidates$}
        \State $ddg \gets \Call{BuildInDDG}{rdo}$
        \State $dataStringSet \gets \Call{StringfyInDDG}{ddg}$
        \State $patternIndexes \gets \Call{DDGAutomaton.run}{dataStringSet}$
        \ForEach{$i \in patternIndexes$}
            \State $matches[i].append(rdo)$
        \EndFor
    \EndFor
    
\\
    \LineComment Return a list with the matches of each pattern
    \State \Return matches
\EndFunction
\label{alg:ddg}
\end{algorithmic}
\end{algorithm}

\subsection{CDG Matching}
\label{sec:cdg}

As discussed before, the control structure of the generic MLIR representation is organized as a hierarchical representation of operations, regions, and basic blocks. A region in MLIR is defined by a control-flow operation called  \textit{Region Defining Operation} (RDO), and each region may contain RDOs defining nested regions. In \rlst{firpattern}, for example, the first \texttt{fir.if} operation (line 7) is an RDO that defines two regions: (a) in the first region are the operations to be performed if the condition is true (lines 8--9); and (b) in the second, the operations performed if it is false (lines 11--22). Region (b) also has a nested \texttt{fir.if} (RDO) (line 16) defining two other regions nested in (b). 
Overall, the CDG matching algorithm works as shown in \ralg{cdg}. First, \textsc{CDGMatch} takes the MLIR code for the input and idiom patterns  (line 1). It then  traverses the MLIR code of each idiom pattern (\texttt{pat}), building a string representation of the sequence of RDOs that define the pattern's MLIR control-flow regions (lines 5 -- 10). This way, the various idiom  CDGs are stored as a set of strings (\texttt{patStrings}) where each string represents a sequence of control-flow regions for one idiom MLIR code. We call these strings \textit{control-strings}. Each control-string originates in an RDO operation inside the  MLIR code.

\begin{minipage}{0.24\textwidth}
\begin{lstlisting}[language={[90]Fortran}, label={lst:idiom-f90}, caption={Fortran idiom  code.}]
subroutine sum(test, val)
  integer :: val, test

  IF (test == 1) THEN
    val = 1
  ELSE
    val = val - 1
    IF (val == 1) THEN
      test = 0
    END IF
  END IF
  
end subroutine
\end{lstlisting}
\end{minipage}
\hspace{5mm}
\begin{minipage}{0.18\textwidth}
\begin{lstlisting}[language={C++}, label={lst:idiom-cdg}, caption={Indented idiom control-string.}]
"fir.if" {
  SEQ
} {
  SEQ
  "fir.if" {
    SEQ
  } {
    SEQ
  }
  SEQ
}
\end{lstlisting}
\end{minipage}

The set of idiom control-strings  (\texttt{patStrings}) is then fed to a method (line 11) that builds the \textsc{CDGAutomaton}. This method allows merging common control-string prefixes to minimize the size of the automaton while marking the states associated to the end of each control-string  final, similarly as shown in \rfig{twigautomaton}. By doing so, the \textsc{CDGAutomaton} works as a compressed representation of the  CDGs of all idiom patterns and can be used later to search idioms inside the input program.

After the CDG automaton is constructed, a similar approach is also performed to build the CDG for the input MLIR code (line 14). The result of this step is a set \texttt{inStrings} that contains the control-strings  associated with fragments of the input program. The set of input control-strings (\texttt{inStrings}) is then fed to function \textsc{CDGAutomaton.Run} to search for those input code fragments that have a similar control-flow structure as an idiom encoded by \textsc{CDGAutomaton} (lines 17 -- 22). An input code fragment is said to \textit{control-match} an idiom if all elements of its control-string take the exact same sequence of states in the CDG automaton as the pattern idiom and end in a final state. 

The call to \textsc{CDGAutomaton.Run(inString)} (line 19) results in a list of pattern indexes (\texttt{patternIndexes}) that identifies which patterns match the control-string \texttt{inString}. The RDO that defines the given \texttt{inString} is retrieved (line 20), and if it matches at least one pattern (line 21), it is inserted as a match candidate (line 22). Finally, \textsc{CDGMatch} returns all input fragments (\texttt{Candidates}) that control-match an idiom pattern. 

In order to ease the understanding of \ralg{cdg}, please consider the Fortran idiom of \rlst{idiom-f90} and its corresponding  CDG control-string (indented)  representation (\rlst{idiom-cdg}). There, curly brackets delimit regions, \texttt{SEQ} are non-empty sequences of non-RDO MLIR operations, and strings (e.g. \texttt{"fir.if"}) are RDOs.

\noindent   Notice that the wrapper function  in \rlst{idiom-f90} (lines 1--2) is not a part of the matching pattern. As aforementioned, it acts only as an interface between input, pattern, and replacement. Therefore, the control-string starts on the first RDO of \rlst{idiom-cdg} (line 1), that  corresponds to the first \texttt{if-else} clause of \rlst{idiom-f90} (line 4). Such RDO is considered to be the RDO that defines the CDG string, which is retrieved at line 20 of \ralg{cdg}.

\subsection{DDG Matching}
\label{sec:ddg}

The output of the CDG pass (\texttt{Candidates}) is a set containing the input code fragments (RDOs) that have a control match with at least one idiom pattern. SMR then proceeds to select from \textit{Candidates} those that have a \textit{Data Dependency Graph} (DDG) match to some idiom. If this happens, that idiom is said to be \textit{matched}.

\begin{lstlisting}[float=tp,language={mlir}, style={mlir}, basicstyle=\tiny, label={lst:firpattern}, caption={MLIR FIR code of idiom in \rlst{idiom-f90}.}]
"func"() ( {
^bb0(%arg0: !fir.ref<i32>, %arg1: !fir.ref<i32>):
  %c1_i32 = "std.constant"() {value = 1 : i32} : () -> i32
  %c0_i32 = "std.constant"() {value = 0 : i32} : () -> i32
  %0 = "fir.load"(%arg0) : (!fir.ref<i32>) -> i32
  %1 = "std.cmpi"(%0, %c1_i32) {predicate = 0 : i64} : (i32, i32) -> i1
  "fir.if"(%1) ( {
    "fir.store"(%c1_i32, %arg1) : (i32, !fir.ref<i32>) -> ()
    "fir.result"() : () -> ()
  },  {
    %2 = "fir.load"(%arg1) : (!fir.ref<i32>) -> i32
    %3 = "std.subi"(%2, %c1_i32) : (i32, i32) -> i32
    "fir.store"(%3, %arg1) : (i32, !fir.ref<i32>) -> ()
    %4 = "fir.load"(%arg1) : (!fir.ref<i32>) -> i32
    %5 = "std.cmpi"(%4, %c1_i32) {predicate = 0 : i64} : (i32, i32) -> i1
    "fir.if"(%5) ( {
      "fir.store"(%c0_i32, %arg0) : (i32, !fir.ref<i32>) -> ()
      "fir.result"() : () -> ()
    },  {
      "fir.result"() : () -> ()
    }) : (i1) -> ()
    "fir.result"() : () -> ()
  }) : (i1) -> ()
  "std.return"() : () -> ()
}) {sym_name = "_QPsum", type = (!fir.ref<i32>, !fir.ref<i32>) -> ()} : () -> ()
\end{lstlisting}

This task is performed by \ralg{ddg} which takes as input the idiom \textit{Patterns} and input \textit{Candidates} (\ralg{cdg} output). The goal of \textsc{DDGMatch}  is to model the input code fragments in \textit{Candidates} and the idioms in \textit{Patterns} as DDGs (\textit{Data Dependency Graphs}) and verify if they are isomorphic. Input and idiom DDGs are built from their generic MLIR representations using a combination of ud-chains and regions. Also, relevant particularities of a dialect are encoded in the strings through SMR's dialect-wise integration, allowing important attributes to be matched, and irrelevant ones, ignored.

\ralg{ddg} starts by calling \textsc{BuildPatDDG}  to build the DDG for every idiom pattern (line 9). Each idiom DDG  is encoded as a set of strings (\texttt{dataStringSet}) by the function \textsc{StringfyPatDDG} (line 10). The strings in \texttt{dataStringSet} are linearized representations of data-flow paths in the DDG, and are called \texttt{data-strings}. The \texttt{dataStringSet} of each and all patterns are then stored together into \texttt{patStringSets} (line 11) so they can later be used to build the  \textsc{DDGAutomaton} (line 14).  The DDG Automaton encodes all the \texttt{patStringSets} data-strings, representing the DDG paths of all idioms' patterns. 

\begin{figure}[!t]
    \centering
    \includegraphics[width=0.35\textwidth,keepaspectratio]{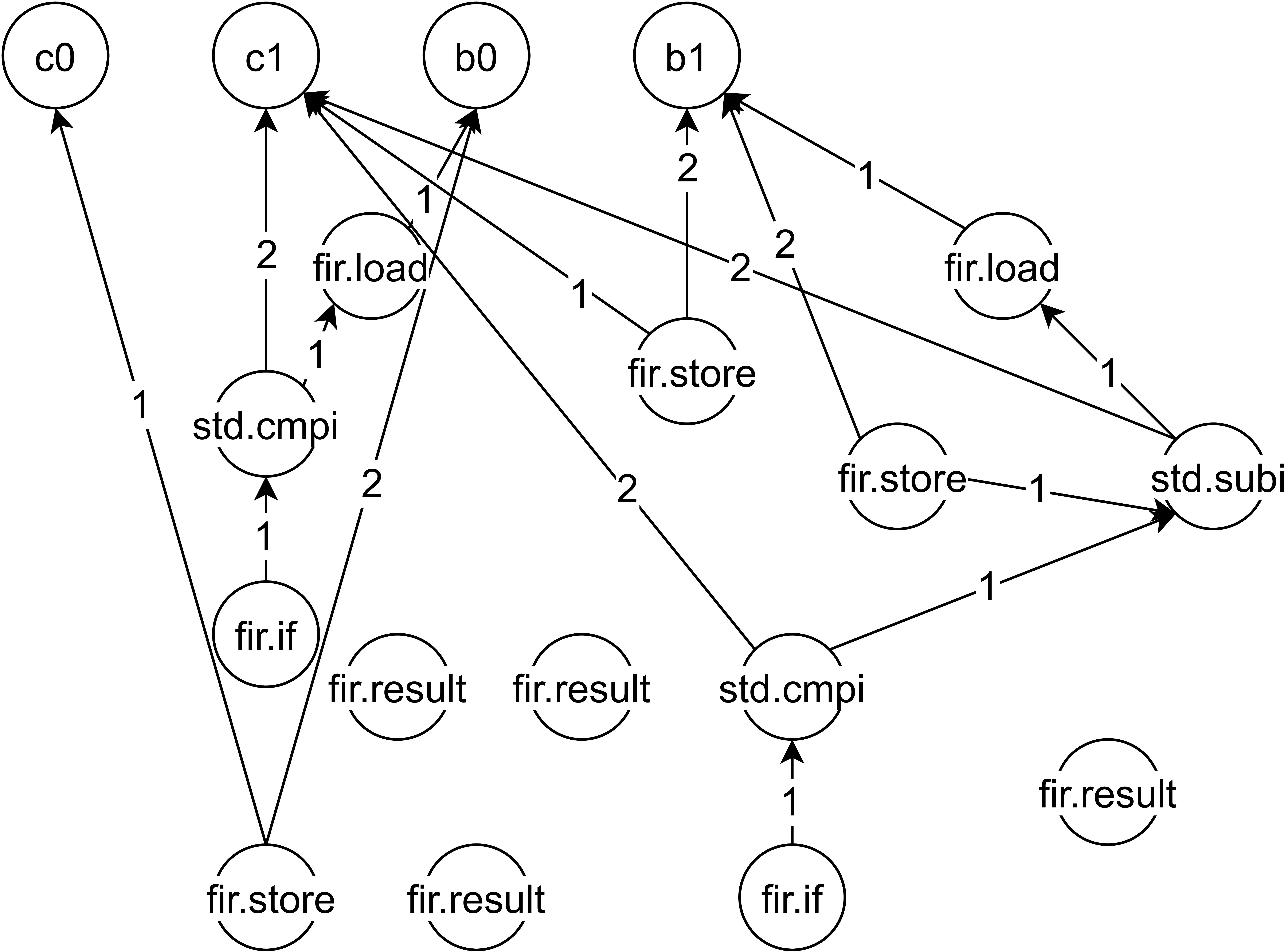}
    \caption{Building ud-chains DDG.}
    \label{fig:udcDag}
\end{figure}

\begin{figure}[!h]
    \centering
    \includegraphics[width=0.35\textwidth,keepaspectratio]{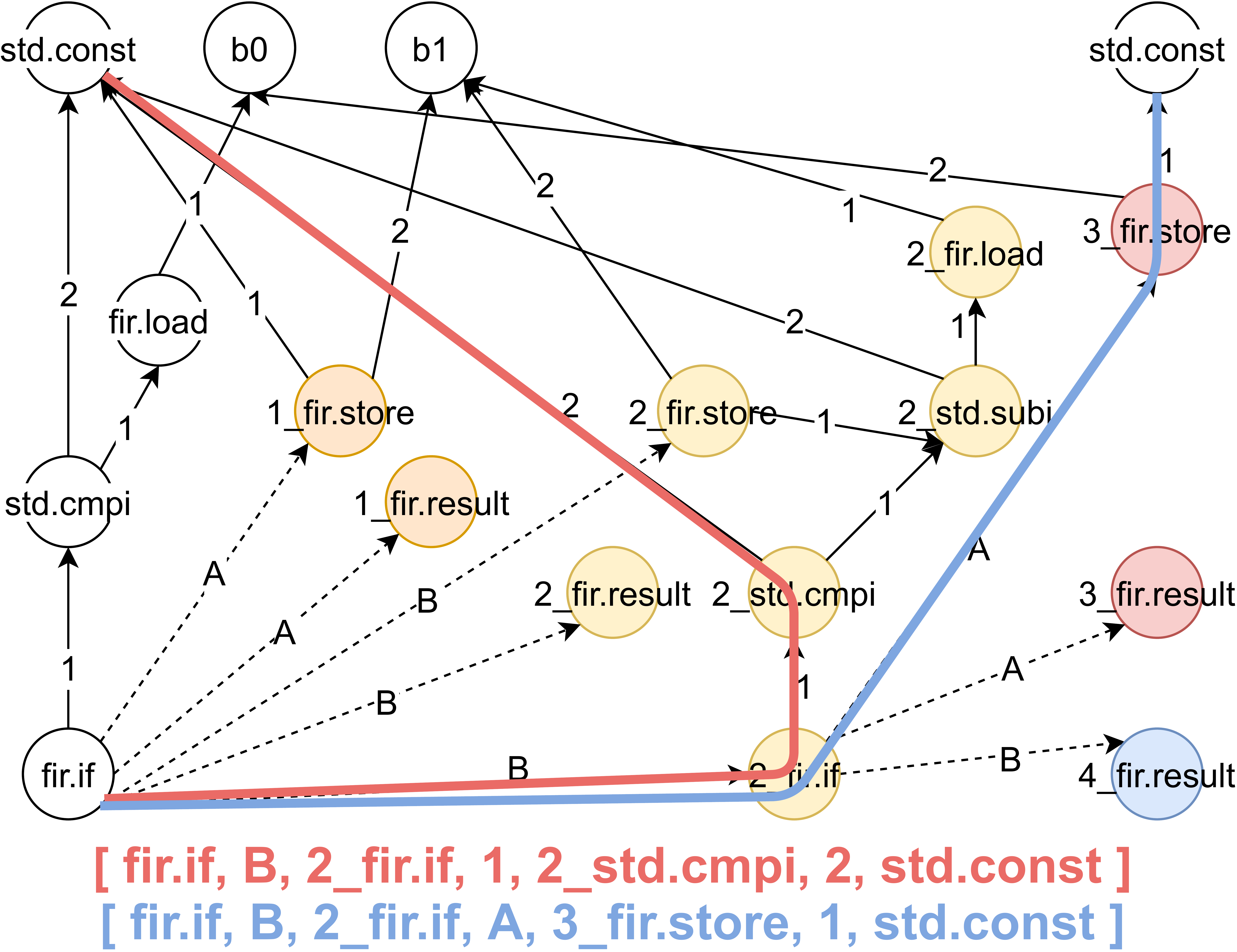}
    \caption{Coloring MLIR regions, adding root, and extracting data-strings.}
    \label{fig:strset}
\end{figure}

After building the \textsc{DDGAutomaton}, each input code  fragment in \texttt{Candidates} is traversed starting at their root RDOs (line 18). Their corresponding DDGs are constructed (line 19), and the associated data-strings are generated (line 20). These strings are fed to the \textsc{DDGAutomaton} to check for a match against all patterns (line 21). The automaton returns all pattern's idioms indexes (\texttt{patternIndexes}) matched by the given input candidate (line 21). If an input fragment matches the i-th pattern, then the root RDO of the input fragment is stored in \texttt{matches[i]} (line 23). Finally, \texttt{DDGMatch}  returns \texttt{matches}, which holds all input MLIR fragments that match a idiom pattern. For example, the i-th idiom matches all input code fragments contained in \texttt{matches[i]}.

Two  methods are used to build DDGs: \texttt{BuildPatDDG} (line 9) and \texttt{BuildInDDG} (line 19). This separation is required due to a difference between the input and pattern DDGs. The wrapper function input arguments in a pattern idiom act as wildcards, not as regular IR values. A wildcard is a node that can match any operation, as long as this operation produces a result that has the same type as the function argument, allowing us to separate where the input ends and the pattern begins in the match.

Although \ralg{ddg} seems  simple, it hides a complex set of tasks that considerably extends the approach originally proposed by Aho et al. in~\cite{Aho1989}. These tasks are performed by two key functions that are described in detail below: (a) DDG Building (lines 9 and 19); and (b) DDG Stringfying (lines 10 and 20).

\subsubsection{Building DDG}

One of the advantages of operating  SMR  on top of a generic MLIR representation is that it can be interpreted directly as a \textit{Rooted Directed Acyclic Graph} (RDAG) by using the MLIR regions and \textit{ud-chains}. 

The idiom in \rlst{idiom-f90} and its corresponding  MLIR representation in \rlst{firpattern} are used here again to show the workings of \ralg{ddg}. To illustrate that, please refer to \rfigs{udcDag}{strset}, which shows the steps required to build the DDG of the MLIR code in \rlst{firpattern}.

From the ud-chains shown in \rlst{firpattern}, it is easy to build a DAG that  serves as the basis for the final DDG. Each MLIR operation corresponds to a node in the graph, and it is labeled by a string composed of the name and relevant attributes of the operation. Note that this identification is not unique: there may be different nodes with the same string label. To connect these nodes, we convert the MLIR ud-chains to directed edges, where the source is the MLIR operation that uses a variable, and the destination is the operation that defines that variable. Since there are operations in which the order of arguments matters (subtraction, for example), we also enumerate the outgoing edges of an operation according to the index of the input operand: the edge corresponding to the first operand is enumerated 1, the edge of the second is 2, etc. This representation is illustrated by the graph in \rfig{udcDag}. By observing that graph, one can  notice two problems that prevent it from being used as a representation of the \rlst{firpattern} code: \textit{control-flow incompleteness} and absence of a root.

The MLIR control regions of \rlst{firpattern} are not represented in the graph of \rfig{udcDag}. Therefore, there is a loss of information regarding the control structure of the code in \rlst{firpattern}, making it incomplete. A possible solution for this is to assign a color to each region in \rlst{firpattern} so that all operations (graph nodes) are colored according to the region in which they are located. To represent this coloring, we identify each color with an integer ID prefixed on the string representation of the node. For example, if a node is in region 2 (yellow), e.g. \texttt{std$\_$cmpi}, it will be labeled  \texttt{2$\_$std.cmpi}. This modification allows us to group MLIR operations according to their respective regions, preserving control-flow information. Such coloring is exemplified in \rfig{strset}.

By analyzing \rfig{udcDag}, one can notice that there are multiple nodes without incident edges, making the graph disjointed and rootless. Such nodes represent MLIR operations that do not define SSA variables and thus are not destinations of any edges created by the ud-chains in \rfig{udcDag}. Examples of such MLIR operations are \texttt{fir.if}, \texttt{fir.store} and \texttt{fir.result}. From now on, these nodes  will be called  \textit{potential roots}. As described below, two issues need to be addressed for the graph to be rooted.

\textbf{Finding Region Edges:}  First, the disjoint DDG graph components must be connected. To address that, let us define \textit{Region Edges}, which are labeled with capital letters in \rfig{strset}. Regions of an MLIR operation are also ordered, and therefore these edges are created alphabetically: first region has edge A, second, edge B, and so on. A region edge $X \xrightarrow{R} Y$ is created whenever there is a potential root node $Y$ that lies within a region $R$ defined by the RDO $X$. For example, in \rfig{strset} operation \texttt{fir.store} of (line 8 of \rlst{firpattern}) is in region $A$ defined by operation \texttt{fir.if} (line 7 of \rlst{firpattern} ). Thus, in \rfig{strset} \texttt{fir.if}$\xrightarrow{A}$\texttt{fir.store} is a region edge. All region edges in the figure are marked as dashed lines.

\textbf{Identifying DDG Root:} 
By combining ud-chains and region edges, we attain the rooted graph of  \rfig{strset}. Since the wrapper function is not a part of the match, the outer-most \texttt{fir.if} operation will not be linked to its parent wrapper function. Thus, as long as there are no sequential RDOs in the wrapper function's body (which is one of the constraints mentioned \rsec{smr-limitations}), there will be only one root RDO after linking the ud-chains and region edges. In \rfig{strset}, such RDO is the \texttt{fir.if} root.

\vspace{-0.4cm}
\subsubsection{Building the DDG Automaton}

In order to build the \texttt{DDGAutomaton} (line 14), the pattern idioms DDG must be represented as strings. This conversion is done by the \texttt{stringifyPatDDG} method (line 10), which takes a pattern DDG and returns a set of data-strings (i.e. \texttt{patStringSet}) that are stored into \texttt{patStringSets}. After the \texttt{DDGAutomaton} is built, the DDG of \texttt{Candidates}' RDOs are extracted (line 18--19) and are  converted to a set of strings by function \texttt{stringifyInDDG}, producing \texttt{dataStringSet} (line 20), which is then fed to the automaton (line 21) for matching. 

The conversion from a DDG  to \texttt{dataStringSet} is trivial. Just list all possible paths from the root to the leaves of the DDG, so that each path  is composed by the concatenation of the identifiers labeled at the nodes and edges. This process is illustrated in \rfig{strset}. The blue path  data-string shows the data-dependency execution path that starts at the region defined by the root \texttt{fir.if} operation in line 7 of \rlst{firpattern} (line numbers from now on refere to that listing). From there, the path follows to the \texttt{fir.if} operation (line 16)  that works as an RDO of an inner (yellow) region in \rfig{strset}. It then proceeds to operation \texttt{std.cmpi} (line 15) which uses constant in \texttt{std.const \%c1\_i32} (line 3). On the other hand, the red path (i.e. data-string) starts at the same root \texttt{fir.if} (line 7), also continues to the inner region \texttt{fir.if} (line 16), but it then diverges through region edge A, reaching the \texttt{3$\_$fir.store} operation (line 17) of the pink region, which uses \texttt{std.const \%c0\_i32} (line 4). Applying this process to every path of an idiom pattern DDG will result in a set of data-strings (\texttt{dataStringSet}) representing the pattern.

\subsection{Idiom Re-writing}
\label{sec:rewriting}

The result of an idiom matching returns a bijection between an idiom  and fragments of the input code. This bijection makes the task of idiom re-writing very simple. SMR needs only to remove from the MLIR code all the operations that can be reached from  the root of the DDG of \rfig{strset}, and then replace the DDG root operation with a  call to the wrapper function of the idiom replacement code in the PAT description. In \rlst{example} the dot product C idiom  is replaced  by a call to its replacement code which  eventually  calls \texttt{cblas$\_$sdot} from the CBLAS library (line 9).

\vspace{-0.2cm}
\subsection{SMR Limitations}
\label{sec:smr-limitations}
This first version of SMR was designed to handle only the cases of idioms that are reducible CFGs as defined in~\cite{Lam2007}. Moreover, a set of additional constraints have  been adopted to accelerate the design of SMR: (a) idiom patterns  may not have sequential RDOs in the wrapper function's body (only nested RDOs). For example,  if an input idiom is composed of two sequential (or non-nested) loops, it cannot be matched using a single pattern idiom.  This does not prevent the user from matching the input partially by writing two pattern idioms, one for each loop; (b) Regions must contain exactly one basic block, otherwise, the DDG would have to model the possible paths between these blocks; (d) Operations must define at most one result operand. To allow more than one result would require each edge to encode not only the input operand position but also the position of the output operand being used; (e) Variadic operations are not supported. Removing such constraints is possible, but to speed up the design of the initial version of SMR, this was left as future work.

\vspace{-0.2cm}
\subsection{FIR vs. CIL} 
As aforementioned, the lowering process from source code to MLIR  depends on the choices of the dialect designer. Hence, similar semantic clauses can be lowered in different ways. An example of such is the difference between CIL and FIR regarding the lowering of \texttt{if-else} clauses. While FIR directly handles the clause with regions, CIL uses the traditional basic blocks method with branching comparison operations, generating a lower level MLIR representation than FIR. That said, FIR can still fall into the same scenario when unstructured code or branching operations are in the input code. If a Fortran \texttt{EXIT} statement is used within a loop, for example, the IR generated by such loop falls directly into an MLIR basic block representation. The same happens when dealing with \texttt{SELECT CASE} statements. In most other aspects, CIL and FIR are quite similar. However, while FIR is currently well supported by the community, CIL has not received any official contributions since it was presented to the LLVM community, rendering it a more crude MLIR dialect when compared to FIR. Regardless, at the writing of this paper, and to the best of our knowledge, CIL is the only available MLIR dialect representation for C/C++.

\vspace{-0.2cm}
\section{Related Work}
\label{sec:related_works}
Pattern matching techniques have  been used for decades to implement compiler optimizations. While early research mainly focused on IR pattern matching for code generation~\cite{Aho1976,Pelegri-Llopart1988,Aho1989,Fraser1992}, recent works explored more complex patterns that include program control flow~\cite{Ginsbach2018, Carvalho2021, Lucke2021, Chelini2021}. LLVM already provides a pattern matching mechanism, which has been used in ~\cite{Carvalho2021} to replace  computation kernels such as GEMM with optimized implementations. Although their approach results in  good speedups, it is not  very flexible.  Every new \textit{kernel} (i.e., idiom) they introduce needs to be  hard-coded into the compiler. MLIR offers a more advanced and generic pattern matching and rewriting tool~\cite{Lattner2021} based on a specific MLIR dialect known as the \textit{Pattern Descriptor Language} (PDL) and a DAG Rewriter mechanism.  PDL provides a declarative description of IR patterns and their replacements while also using the generic representation of MLIR to enable that to other MLIR dialects. PDL works quite well  in the case of simple patterns, but it is not yet possible to use it to describe more complex idioms such as matrix multiplications~\cite{Riddle2021}. In fact, the generic representation of MLIR is very low-level, and algorithms relying on advanced control structures are complex to express in it. On the other hand, dedicated MLIR dialects can be used to capture idioms for optimization. For example, Uday Bondhugula~\cite{Bondhugula2020} has matched GEMM idioms and replaced them with BLAS calls to improve performance. But similarly, as in~\cite{Carvalho2021} matching/replacement is hard-coded inside the compiler.

Several works try to expand LLVM and MLIR pattern matching through domain-specific languages to ease the matching of complex idioms by using custom functional languages~\cite{Lucke2021, Chelini2021} or a constraint-based language~\cite{Ginsbach2018}. For all those approaches, the idiom descriptions are synthesized as LLVM IR or MLIR passes that perform  pattern matching. The fundamental difference between those works is that Chelini and al. use pattern matching to raise the abstraction level of the intermediate representation of general-purpose languages to  allow domain-specific optimizations,  while L\"{u}cke and al. mostly focus on simplifying pattern matching for a domain-specific IR. On the performance side, Chelini and al. already demonstrate their approach  for linear algebra operations against standard compilation optimizations and polyhedral optimizations.  All such approaches clearly allow a more compact way to match and rewrite idioms. Nevertheless, they require re-compiling the compiler and/or are not very friendly to express idioms as source code is in SMR, two undesirable features when targeting idiom exploration. 

Barthels and al.~\cite{Barthels2020} explore the automatic rewriting of linear algebra problems from high-level representations (Julia, Eigen, or Matlab) using pattern matching and rewriting. Their work uses the mathematical properties of linear algebra operations and data structures to derive efficient implementations. Although they clearly demonstrate the potential of raising, their approach only works when the target program uses  high-level function calls to algebraic operations.  

Similar to SMR, the XARK compiler~\cite{Arenaz2008} uses a two-phase  approach to match idioms. However, SMR uses automaton string matching, which is more efficient than matching an IR graph representation directly, and starts by analyzing the CDG to quickly eliminate non-relevant idioms. Additionally, XARK reduces idioms expressiveness more than SMR since it requires linearizing the accesses to multidimensional arrays.

Polly~\cite{Grosser2012} focuses on polyhedral loop transformations, but also uses pattern matching in some cases. Polygeist~\cite{Moses2021} is a C front-end for the MLIR Affine dialect which was not released in time to be used by this work.

Verified Lifting~\cite{Kamil2016} uses a formal verification approach to perform pattern matching and higher-level rewriting. Their approach showed good results but it is specialized to stencil codes and, in some cases, need to include user annotations to ensure the correct rewrite.

\vspace{-0.2cm}
\section{Experimental Results}
\label{sec:results}

This paper aims to propose and validate SMR, a programmer-friendly approach for idiom matching and re\-writing. Contrary to other works~\cite{Chelini2021}\cite{Bondhugula2020}, we do not seek to evaluate the performance of the rewritten programs for other established polyhedral-based techniques, as this has already been explored~\cite{Chelini2021}.  With this in mind, three sets of experiments have been designed to evaluate SMR. First,  a set of Fortran programs was used to validate the correctness of the rewritten code by demonstrating the expected speedups. Fortran was selected because its corresponding MLIR dialect (FIR) is quite stable and well maintained by the community, while C's dialect (CIL) is still in a very brittle state. In the second set of experiments, the impact  of SMR matching/replacement on the standard FIR compilation time was measured.  Finally, the goal of the last set of experiments was to evaluate the ability of SMR to match another language dialect (C/CIL). Evaluation of SMR's  potential to capture more generic patterns was left as future work. 

All experiments were performed using a dual Intel Xeon Silver 4208 CPU @ 2.10GHz with 16 cores total and 191 GiB of RAM running Ubuntu 18.04. As for the software tool-chain, the following commits/versions have been used: (a)  Flang (FIR) commit  \texttt{8abd290} \cite{FIRrepository2021}; (b) CIL commit  \texttt{195acc3} \cite{CILrepository2021}; (c) LLVM/MLIR commit \texttt{1fdec59} \cite{LLVM/MLIRrepository2021}; (d) \texttt{OpenBLAS} version 0.2.20~\cite{Xianyi2020}; and (e) GFortran version \texttt{4.8.5}. Experiments used programs from Fortran Polybench 1.0 benchmark~\cite{Narayan2012}, and a set of 6 large well-known C programs from different application domains  (top labels of \rtab{c-matches}) that perform intense arithmetic computations, extracted using the  Angha tool~\cite{DaSilva2021}. All experiments were executed following up the benchmark execution guidelines. They  used standard reference inputs and were executed 5 times showing small execution time variations (<~5\%).  SMR has detected no false-positive idioms in all experiments.

\begin{figure}[t]
  \includegraphics[width=\linewidth]{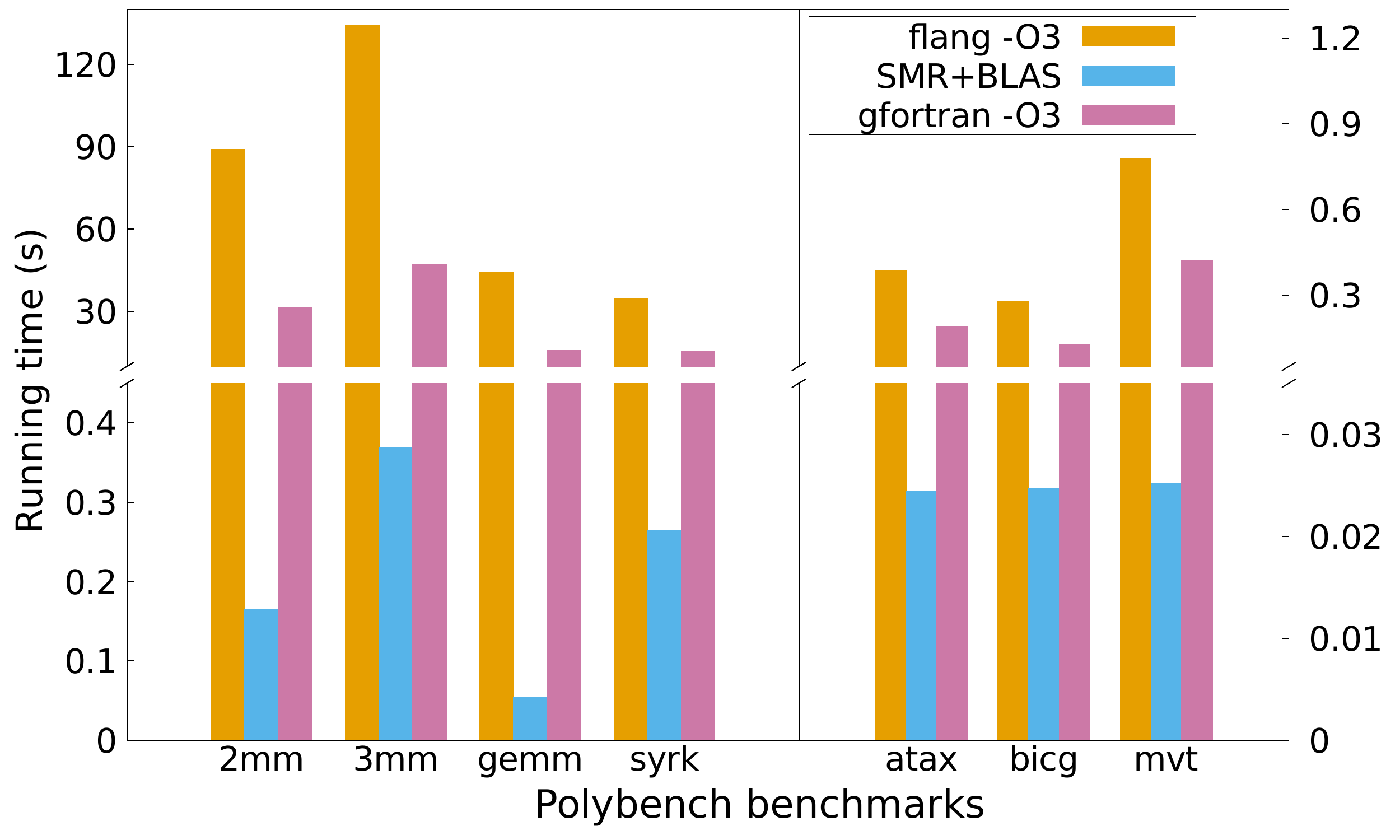}
  \vspace{-4ex}%
  \caption{Polybench running time after BLAS replacement.~\protect\footnotemark}
  \label{fig:res_poly_Fortran}
\end{figure}

In the first experiment, a set of idioms targeting Fortran double-precision BLAS kernels have been designed using the PAT language and applied to Fortran Polybench programs  for matching. Idioms were substituted by the corresponding Fortran BLAS calls, and execution times were measured with the \texttt{LARGE\_DATASET} option (\rfig{res_poly_Fortran}). As shown in the figure, the performance of the Flang version used in this experiment lags behind GFortran. When using the FIR front-end together with SMR idiom detection and Fortran BLAS replacement (SMR+BLAS, blue bar), all programs showed speed-ups, ranging from 5x to 295x. For some programs, SMR could not replace the kernel with a single BLAS call, either because there is no matching BLAS call for the kernel or because it does not meet SMR restrictions. In such cases, partial replacements by multiple BLAS calls occurred.  This resulted in  speed-ups, as in the case of \texttt{2mm} (replaced by two GEMMs),  \texttt{atax} (partially replaced by two GEMVs) and \texttt{bigc} (partially replaced by GEMVs calls). Other programs from the benchmark were neither fully nor partially rewritten (e.g. \texttt{syr2k}), given they are composed by sequences of RDOs, an SMR restriction, and could not be separated into multiple BLAS calls. As discussed before, this restriction is a solvable issue, and work is underway to address it. 

\footnotetext{Due to the differences in execution times, y-axes were broken, and programs separated into two groups }

\begin{figure}
  \includegraphics[width=\linewidth]{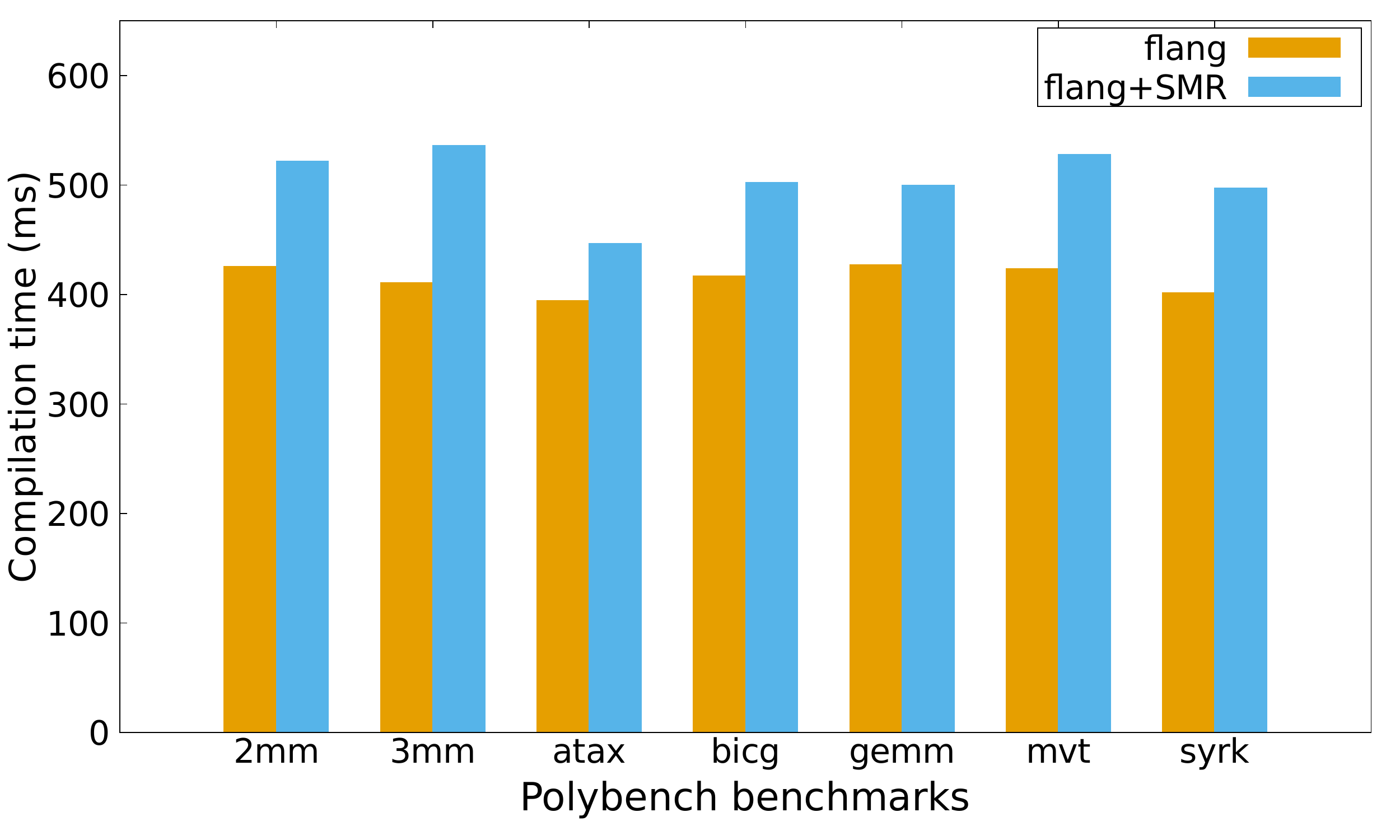}
  \vspace{-4ex}%
  \caption{FIR compilation time with/without SMR+BLAS.}
  \label{fig:compilation_time}
\end{figure}
\begin{table}[]
\resizebox{0.33\textwidth}{!}{%
\begin{tabular}{|l|c|c|c|c|c|c|c|}
\hline
\textbf{Idiom} & \rot{Darknet~\cite{Redmon2016}} & \rot{Cello~\cite{Holden2015}} & \rot{Exploitdb~\cite{OffensiveSecurity2021}} & \rot{Ffmpeg~\cite{FFmpegDevelopers2021}} & \rot{Hpgmg~\cite{Adams2014}} & \rot{Nekrs~\cite{Fischer2021}} & \rot{\textbf{Total}} \\  \hline 
saxpy & 1 & & & & & & 1 \\  \hline
scopy & 1 & & & & & &  1 \\  \hline
sdot & 1 &  &  & 1 & & & 2 \\  \hline
sgemm & 4 & & & & & & 4 \\  \hline
scall & 2 & & & & & & 2 \\  \hline
ddot &  & 1 &  & & 1 & 2 & 4 \\  \hline
dgemm &  & & 1 & & &  3 & 4 \\  \hline
dgemmv &  & &  & & &  1 & 1 \\  \hline
dscal &  & &  & & &  3 & 3 \\  \hline \hline
\textbf{Total} & 9  & 1 & 1 & 1 & 1 & 9 & 22 \\  \hline
\end{tabular}%
}
\caption{Matching with CIL and CBLAS idioms.}
\label{tab:c-matches}
\vspace{-1cm}
\end{table}

In the second set of experiments, execution times for Fortran Polybench programs were  measured using   FIR compilation with and without SMR+BLAS  (\rfig{compilation_time}). As shown in the figure, the compilation overhead ranged from 52 ms to 125 ms. The reader should notice that SMR+BLAS compiles and builds the automata for each set of idioms in a PAT file. This is not required, though. In the future, we intend to compile the PAT file once and save the idioms' automata for reuse,  thus eliminating 60\% -- 85\% of this overhead.

In the final set of experiments, 9 idioms associated with C BLAS calls were  used with the CIL front-end to perform idiom matching on 6 C programs. Program re-writing was skipped due to the unstable state of the CIL front-end.  \rtab{c-matches} shows the number of idiom occurrences detected per input program. Overall, 22 idioms occurrences  were detected in all six programs, with  Darknet and Nekrs matching 9  occurrences each, respectively using 5 and 4 distinct idioms. As a comparison, Polly v13.0.0 was only able to match the \textit{gemm} idiom: 1 occurrence in Darknet and 3 in Nekrs. We also confirmed that PDL was not able to handle those idioms since it does not yet support regions \cite{PDLRegion}.

\section{Conclusions}
\label{sec:conclusions}
This paper proposes SMR, a source-code and MLIR based idiom matching and re-writing approach. By using MLIR generality, SMR could match both Fortran and C programs through their corresponding dialects (FIR and CIL). In FIR's case, it was also able to optimize benchmarks through code rewriting. Future extensions will address the SMR constraints listed above, enable inter-language replacement and matching of novel accelerated ISA extensions and library calls, and replacement with high-level operations from other MLIR dialects.

\bibliographystyle{ACM-Reference-Format}
\bibliography{smr}


\begin{thebibliography}{45}


\ifx \showCODEN    \undefined \def \showCODEN     #1{\unskip}     \fi
\ifx \showDOI      \undefined \def \showDOI       #1{#1}\fi
\ifx \showISBNx    \undefined \def \showISBNx     #1{\unskip}     \fi
\ifx \showISBNxiii \undefined \def \showISBNxiii  #1{\unskip}     \fi
\ifx \showISSN     \undefined \def \showISSN      #1{\unskip}     \fi
\ifx \showLCCN     \undefined \def \showLCCN      #1{\unskip}     \fi
\ifx \shownote     \undefined \def \shownote      #1{#1}          \fi
\ifx \showarticletitle \undefined \def \showarticletitle #1{#1}   \fi
\ifx \showURL      \undefined \def \showURL       {\relax}        \fi
\providecommand\bibfield[2]{#2}
\providecommand\bibinfo[2]{#2}
\providecommand\natexlab[1]{#1}
\providecommand\showeprint[2][]{arXiv:#2}

\bibitem[\protect\citeauthoryear{Adams, Brown, Shalf, {Van Straalen},
  Strohmaier, and Williams}{Adams et~al\mbox{.}}{2014}]%
        {Adams2014}
\bibfield{author}{\bibinfo{person}{Mark~F. Adams}, \bibinfo{person}{Jed Brown},
  \bibinfo{person}{John Shalf}, \bibinfo{person}{Brian {Van Straalen}},
  \bibinfo{person}{Erich Strohmaier}, {and} \bibinfo{person}{Samuel Williams}.}
  \bibinfo{year}{2014}\natexlab{}.
\newblock \bibinfo{booktitle}{\emph{{HPGMG 1.0: A Benchmark for Ranking High
  Performance Computing Systems}}}.
\newblock \bibinfo{type}{{T}echnical {R}eport}. \bibinfo{institution}{LBNL}.
  \bibinfo{pages}{11} pages.
\newblock


\bibitem[\protect\citeauthoryear{Aho, Ganapathi, and Tjiang}{Aho
  et~al\mbox{.}}{1989}]%
        {Aho1989}
\bibfield{author}{\bibinfo{person}{Alfred~V. Aho}, \bibinfo{person}{Mahadevan
  Ganapathi}, {and} \bibinfo{person}{Steven W.~K. Tjiang}.}
  \bibinfo{year}{1989}\natexlab{}.
\newblock \showarticletitle{{Code generation using tree matching and dynamic
  programming}}.
\newblock \bibinfo{journal}{\emph{ACM Transactions on Programming Languages and
  Systems}} \bibinfo{volume}{11}, \bibinfo{number}{4} (\bibinfo{date}{oct}
  \bibinfo{year}{1989}).
\newblock
\showISSN{0164-0925}
\urldef\tempurl%
\url{https://doi.org/10.1145/69558.75700}
\showDOI{\tempurl}


\bibitem[\protect\citeauthoryear{Aho and Johnson}{Aho and Johnson}{1976}]%
        {Aho1976}
\bibfield{author}{\bibinfo{person}{A.~V. Aho} {and} \bibinfo{person}{S.~C.
  Johnson}.} \bibinfo{year}{1976}\natexlab{}.
\newblock \showarticletitle{{Optimal Code Generation for Expression Trees}}.
\newblock \bibinfo{journal}{\emph{J. ACM}} \bibinfo{volume}{23},
  \bibinfo{number}{3} (\bibinfo{date}{jul} \bibinfo{year}{1976}).
\newblock
\showISSN{0004-5411}
\urldef\tempurl%
\url{https://doi.org/10.1145/321958.321970}
\showDOI{\tempurl}


\bibitem[\protect\citeauthoryear{Arenaz, Tourĩo, and Doallo}{Arenaz
  et~al\mbox{.}}{2008}]%
        {Arenaz2008}
\bibfield{author}{\bibinfo{person}{Manuel Arenaz}, \bibinfo{person}{Juan
  Tourĩo}, {and} \bibinfo{person}{Ramon Doallo}.}
  \bibinfo{year}{2008}\natexlab{}.
\newblock \showarticletitle{{XARK: An extensible framework for automatic
  recognition of computational kernels}}.
\newblock \bibinfo{journal}{\emph{ACM Transactions on Programming Languages and
  Systems}} \bibinfo{volume}{30}, \bibinfo{number}{6} (\bibinfo{year}{2008}).
\newblock
\showISSN{01640925}
\urldef\tempurl%
\url{https://doi.org/10.1145/1391956.1391959}
\showDOI{\tempurl}


\bibitem[\protect\citeauthoryear{artifact repository}{artifact
  repository}{2021}]%
        {smr-artifact-repo}
\bibfield{author}{\bibinfo{person}{SMR artifact repository}.}
  \bibinfo{year}{2021}\natexlab{}.
\newblock \bibinfo{title}{{Master branch}}.
\newblock
\newblock
\urldef\tempurl%
\url{https://gitlab.com/parlab/smr-artifact}
\showURL{%
\tempurl}


\bibitem[\protect\citeauthoryear{Barthels, Psarras, and Bientinesi}{Barthels
  et~al\mbox{.}}{2020}]%
        {Barthels2020}
\bibfield{author}{\bibinfo{person}{Henrik Barthels}, \bibinfo{person}{Christos
  Psarras}, {and} \bibinfo{person}{Paolo Bientinesi}.}
  \bibinfo{year}{2020}\natexlab{}.
\newblock \showarticletitle{{Automatic Generation of Efficient Linear Algebra
  Programs}}. In \bibinfo{booktitle}{\emph{Proceedings of the Platform for
  Advanced Scientific Computing Conference, PASC 2020}}.
\newblock
\showISBNx{9781450379939}
\urldef\tempurl%
\url{https://doi.org/10.1145/3394277.3401836}
\showDOI{\tempurl}
\showeprint[arxiv]{arXiv:1912.12924v1}


\bibitem[\protect\citeauthoryear{binary repository}{binary repository}{2021}]%
        {smr-binary-repo}
\bibfield{author}{\bibinfo{person}{SMR binary repository}.}
  \bibinfo{year}{2021}\natexlab{}.
\newblock \bibinfo{title}{{tag cgo}}.
\newblock
\newblock
\urldef\tempurl%
\url{https://gitlab.com/parlab/pat-compiler/-/tree/cgo}
\showURL{%
\tempurl}


\bibitem[\protect\citeauthoryear{Blume, Eigenmann, Faigin, Grout, Hoeflinger,
  Padua, Petersen, Pottenger, Rauchwerger, Tu, and Weatherford}{Blume
  et~al\mbox{.}}{1995}]%
        {Blume1995}
\bibfield{author}{\bibinfo{person}{William Blume}, \bibinfo{person}{Rudolf
  Eigenmann}, \bibinfo{person}{Keith Faigin}, \bibinfo{person}{John Grout},
  \bibinfo{person}{Jay Hoeflinger}, \bibinfo{person}{David Padua},
  \bibinfo{person}{Paul Petersen}, \bibinfo{person}{William Pottenger},
  \bibinfo{person}{Lawrence Rauchwerger}, \bibinfo{person}{Peng Tu}, {and}
  \bibinfo{person}{Stephen Weatherford}.} \bibinfo{year}{1995}\natexlab{}.
\newblock \showarticletitle{{Polaris: Improving the effectiveness of
  parallelizing compilers}}. In \bibinfo{booktitle}{\emph{Lecture Notes in
  Computer Science (including subseries Lecture Notes in Artificial
  Intelligence and Lecture Notes in Bioinformatics)}},
  Vol.~\bibinfo{volume}{892}. \bibinfo{pages}{141--154}.
\newblock
\showISBNx{354058868X}
\showISSN{16113349}
\urldef\tempurl%
\url{https://doi.org/10.1007/bfb0025876}
\showDOI{\tempurl}


\bibitem[\protect\citeauthoryear{Bondhugula}{Bondhugula}{2020}]%
        {Bondhugula2020}
\bibfield{author}{\bibinfo{person}{Uday Bondhugula}.}
  \bibinfo{year}{2020}\natexlab{}.
\newblock \bibinfo{title}{{High Performance Code Generation in MLIR: An Early
  Case Study with GEMM}}.
\newblock , \bibinfo{numpages}{23}~pages.
\newblock
\showeprint[arxiv]{2003.00532}


\bibitem[\protect\citeauthoryear{Carvalho, Kuzma, Korostelev, Amaral, Barton,
  Moreira, and Araujo}{Carvalho et~al\mbox{.}}{2021}]%
        {Carvalho2021}
\bibfield{author}{\bibinfo{person}{Joao P. L.~De Carvalho},
  \bibinfo{person}{Braedy Kuzma}, \bibinfo{person}{Ivan Korostelev},
  \bibinfo{person}{Jos{\'{e}}~Nelson Amaral}, \bibinfo{person}{Christopher
  Barton}, \bibinfo{person}{Jos{\'{e}} Moreira}, {and} \bibinfo{person}{Guido
  Araujo}.} \bibinfo{year}{2021}\natexlab{}.
\newblock \showarticletitle{{KernelFaRer : Replacing Native-Code Idioms with
  High-Performance Library Calls}}.
\newblock \bibinfo{journal}{\emph{ACM Transactions on Architecture and Code
  Optimization}} (\bibinfo{year}{2021}).
\newblock


\bibitem[\protect\citeauthoryear{Chelini, Drebes, Zinenko, Cohen, Vasilache,
  Grosser, and Corporaal}{Chelini et~al\mbox{.}}{2021}]%
        {Chelini2021}
\bibfield{author}{\bibinfo{person}{Lorenzo Chelini}, \bibinfo{person}{Andi
  Drebes}, \bibinfo{person}{Oleksandr Zinenko}, \bibinfo{person}{Albert Cohen},
  \bibinfo{person}{Nicolas Vasilache}, \bibinfo{person}{Tobias Grosser}, {and}
  \bibinfo{person}{Henk Corporaal}.} \bibinfo{year}{2021}\natexlab{}.
\newblock \showarticletitle{{Progressive Raising in Multi-level IR}}. In
  \bibinfo{booktitle}{\emph{CGO 2021 - Proceedings of the 2021 IEEE/ACM
  International Symposium on Code Generation and Optimization}}.
  \bibinfo{pages}{15--26}.
\newblock
\showISBNx{9781728186139}
\urldef\tempurl%
\url{https://doi.org/10.1109/CGO51591.2021.9370332}
\showDOI{\tempurl}


\bibitem[\protect\citeauthoryear{{Compiler Tree Technologies}}{{Compiler Tree
  Technologies}}{2021}]%
        {cil}
\bibfield{author}{\bibinfo{person}{{Compiler Tree Technologies}}.}
  \bibinfo{year}{2021}\natexlab{}.
\newblock \bibinfo{title}{{CIL}}.
\newblock
\newblock
\urldef\tempurl%
\url{https://github.com/compiler-tree-technologies/cil}
\showURL{%
\tempurl}


\bibitem[\protect\citeauthoryear{Cooper and Torczon}{Cooper and
  Torczon}{2012}]%
        {Cooper2012}
\bibfield{author}{\bibinfo{person}{Keith Cooper} {and} \bibinfo{person}{Linda
  Torczon}.} \bibinfo{year}{2012}\natexlab{}.
\newblock \bibinfo{booktitle}{\emph{{Engineering a Compiler}}}.
\newblock
\showISBNx{9780120884780}
\urldef\tempurl%
\url{https://doi.org/10.1016/c2009-0-27982-7}
\showDOI{\tempurl}


\bibitem[\protect\citeauthoryear{{Da Silva}, Kind, {De Souza Magalhaes}, Rocha,
  {Ferreira Guimaraes}, and {Quinao Pereira}}{{Da Silva} et~al\mbox{.}}{2021}]%
        {DaSilva2021}
\bibfield{author}{\bibinfo{person}{Anderson~Faustino {Da Silva}},
  \bibinfo{person}{Bruno~Conde Kind}, \bibinfo{person}{Jose~Wesley {De Souza
  Magalhaes}}, \bibinfo{person}{Jeronimo~Nunes Rocha},
  \bibinfo{person}{Breno~Campos {Ferreira Guimaraes}}, {and}
  \bibinfo{person}{Fernando~Magno {Quinao Pereira}}.}
  \bibinfo{year}{2021}\natexlab{}.
\newblock \showarticletitle{{ANGHABENCH: A Suite with One Million Compilable C
  Benchmarks for Code-Size Reduction}}. In \bibinfo{booktitle}{\emph{CGO 2021 -
  Proceedings of the 2021 IEEE/ACM International Symposium on Code Generation
  and Optimization}}. \bibinfo{pages}{378--390}.
\newblock
\showISBNx{9781728186139}
\urldef\tempurl%
\url{https://doi.org/10.1109/CGO51591.2021.9370322}
\showDOI{\tempurl}


\bibitem[\protect\citeauthoryear{Documentation.}{Documentation.}{2021}]%
        {llvm-hardware}
\bibfield{author}{\bibinfo{person}{LLVM Documentation.}}
  \bibinfo{year}{2021}\natexlab{}.
\newblock \bibinfo{title}{{Supported hardware.}}
\newblock
\newblock
\urldef\tempurl%
\url{https://llvm.org/docs/GettingStarted.html#hardware}
\showURL{%
\tempurl}


\bibitem[\protect\citeauthoryear{{FFmpeg Developers}}{{FFmpeg
  Developers}}{2021}]%
        {FFmpegDevelopers2021}
\bibfield{author}{\bibinfo{person}{{FFmpeg Developers}}.}
  \bibinfo{year}{2021}\natexlab{}.
\newblock \bibinfo{title}{{FFmpeg tool}}.
\newblock
\newblock
\urldef\tempurl%
\url{https://ffmpeg.org/}
\showURL{%
\tempurl}


\bibitem[\protect\citeauthoryear{Fischer, Kerkemeier, Min, Lan, Phillips,
  Rathnayake, Merzari, Tomboulides, Karakus, Chalmers, and Warburton}{Fischer
  et~al\mbox{.}}{2021}]%
        {Fischer2021}
\bibfield{author}{\bibinfo{person}{Paul Fischer}, \bibinfo{person}{Stefan
  Kerkemeier}, \bibinfo{person}{Misun Min}, \bibinfo{person}{Yu-Hsiang Lan},
  \bibinfo{person}{Malachi Phillips}, \bibinfo{person}{Thilina Rathnayake},
  \bibinfo{person}{Elia Merzari}, \bibinfo{person}{Ananias Tomboulides},
  \bibinfo{person}{Ali Karakus}, \bibinfo{person}{Noel Chalmers}, {and}
  \bibinfo{person}{Tim Warburton}.} \bibinfo{year}{2021}\natexlab{}.
\newblock \bibinfo{booktitle}{\emph{{NekRS, a GPU-Accelerated Spectral Element
  Navier-Stokes Solver}}}.
\newblock \bibinfo{type}{{T}echnical {R}eport}.
\newblock
\showeprint[arxiv]{2104.05829}


\bibitem[\protect\citeauthoryear{{Flang Compiler}}{{Flang Compiler}}{2019}]%
        {fir}
\bibfield{author}{\bibinfo{person}{{Flang Compiler}}.}
  \bibinfo{year}{2019}\natexlab{}.
\newblock \bibinfo{title}{{F18 LLVM Project (fir-dev branch)}}.
\newblock
\newblock
\urldef\tempurl%
\url{https://github.com/flang-compiler/f18-llvm-project/tree/fir-dev}
\showURL{%
\tempurl}


\bibitem[\protect\citeauthoryear{Fraser, Hanson, and Proebsting}{Fraser
  et~al\mbox{.}}{1992}]%
        {Fraser1992}
\bibfield{author}{\bibinfo{person}{Christopher~W. Fraser},
  \bibinfo{person}{David~R. Hanson}, {and} \bibinfo{person}{Todd~A.
  Proebsting}.} \bibinfo{year}{1992}\natexlab{}.
\newblock \showarticletitle{{Engineering a Simple, Efficient Code-Generator
  Generator}}.
\newblock \bibinfo{journal}{\emph{ACM Letters on Programming Languages and
  Systems (LOPLAS)}} \bibinfo{volume}{1}, \bibinfo{number}{3}
  (\bibinfo{year}{1992}), \bibinfo{pages}{213--226}.
\newblock
\showISSN{15577384}
\urldef\tempurl%
\url{https://doi.org/10.1145/151640.151642}
\showDOI{\tempurl}


\bibitem[\protect\citeauthoryear{Ginsbach, Collie, and O'Boyle}{Ginsbach
  et~al\mbox{.}}{2020}]%
        {Ginsbach2020}
\bibfield{author}{\bibinfo{person}{Philip Ginsbach}, \bibinfo{person}{Bruce
  Collie}, {and} \bibinfo{person}{Michael~F.P. O'Boyle}.}
  \bibinfo{year}{2020}\natexlab{}.
\newblock \showarticletitle{{Automatically harnessing sparse acceleration}}. In
  \bibinfo{booktitle}{\emph{CC 2020 - Proceedings of the 29th International
  Conference on Compiler Construction}}. \bibinfo{pages}{179--190}.
\newblock
\showISBNx{9781450371209}
\urldef\tempurl%
\url{https://doi.org/10.1145/3377555.3377893}
\showDOI{\tempurl}
\showeprint[arxiv]{2001.07938}


\bibitem[\protect\citeauthoryear{Ginsbach, Remmelg, Steuwer, Bodin, Dubach, and
  O'Boyle}{Ginsbach et~al\mbox{.}}{2018}]%
        {Ginsbach2018}
\bibfield{author}{\bibinfo{person}{Philip Ginsbach}, \bibinfo{person}{Toomas
  Remmelg}, \bibinfo{person}{Michel Steuwer}, \bibinfo{person}{Bruno Bodin},
  \bibinfo{person}{Christophe Dubach}, {and} \bibinfo{person}{Michael~F.P.
  O'Boyle}.} \bibinfo{year}{2018}\natexlab{}.
\newblock \showarticletitle{{Automatic matching of legacy code to heterogeneous
  APIs: An idiomatic approach}}.
\newblock \bibinfo{journal}{\emph{ACM SIGPLAN Notices}} \bibinfo{volume}{53},
  \bibinfo{number}{2} (\bibinfo{year}{2018}), \bibinfo{pages}{139--153}.
\newblock
\showISBNx{9781450349116}
\showISSN{15232867}
\urldef\tempurl%
\url{https://doi.org/10.1145/3173162.3173182}
\showDOI{\tempurl}


\bibitem[\protect\citeauthoryear{Google}{Google}{2020}]%
        {mlir-docs}
\bibfield{author}{\bibinfo{person}{Google}.} \bibinfo{year}{2020}\natexlab{}.
\newblock \bibinfo{title}{MLIR Documentation}.
\newblock
\newblock
\urldef\tempurl%
\url{https://mlir.llvm.org/docs/}
\showURL{%
\tempurl}


\bibitem[\protect\citeauthoryear{Grosser, Groesslinger, and Lengauer}{Grosser
  et~al\mbox{.}}{2012}]%
        {Grosser2012}
\bibfield{author}{\bibinfo{person}{Tobias Grosser}, \bibinfo{person}{Armin
  Groesslinger}, {and} \bibinfo{person}{Christian Lengauer}.}
  \bibinfo{year}{2012}\natexlab{}.
\newblock \showarticletitle{Polly - Performing polyhedral optimizations on a
  low-level intermediate representation}.
\newblock \bibinfo{journal}{\emph{Parallel Processing Letters}}
  \bibinfo{volume}{22} (\bibinfo{date}{12} \bibinfo{year}{2012}),
  \bibinfo{pages}{1250010}.
\newblock
Issue 04.
\showISSN{0129-6264}
\urldef\tempurl%
\url{https://doi.org/10.1142/S0129626412500107}
\showDOI{\tempurl}


\bibitem[\protect\citeauthoryear{Hoffmann and O'Donnell}{Hoffmann and
  O'Donnell}{1982}]%
        {Hoffmann1982}
\bibfield{author}{\bibinfo{person}{Christoph~M. Hoffmann} {and}
  \bibinfo{person}{Michael~J. O'Donnell}.} \bibinfo{year}{1982}\natexlab{}.
\newblock \showarticletitle{{Pattern Matching in Trees}}.
\newblock \bibinfo{journal}{\emph{Journal of the ACM (JACM)}}
  \bibinfo{volume}{29}, \bibinfo{number}{1} (\bibinfo{year}{1982}),
  \bibinfo{pages}{68--95}.
\newblock
\showISSN{1557735X}
\urldef\tempurl%
\url{https://doi.org/10.1145/322290.322295}
\showDOI{\tempurl}


\bibitem[\protect\citeauthoryear{Holden}{Holden}{2015}]%
        {Holden2015}
\bibfield{author}{\bibinfo{person}{Daniel Holden}.}
  \bibinfo{year}{2015}\natexlab{}.
\newblock \bibinfo{title}{{Cello: Higher level programming in C}}.
\newblock
\newblock
\urldef\tempurl%
\url{https://libcello.org/}
\showURL{%
\tempurl}


\bibitem[\protect\citeauthoryear{Kamil, Cheung, Itzhaky, and
  Solar-Lezama}{Kamil et~al\mbox{.}}{2016}]%
        {Kamil2016}
\bibfield{author}{\bibinfo{person}{Shoaib Kamil}, \bibinfo{person}{Alvin
  Cheung}, \bibinfo{person}{Shachar Itzhaky}, {and} \bibinfo{person}{Armando
  Solar-Lezama}.} \bibinfo{year}{2016}\natexlab{}.
\newblock \showarticletitle{Verified lifting of stencil computations}.
\newblock \bibinfo{journal}{\emph{Proceedings of the 37th ACM SIGPLAN
  Conference on Programming Language Design and Implementation}},
  \bibinfo{pages}{711--726}.
\newblock
\showISBNx{9781450342612}
\urldef\tempurl%
\url{https://doi.org/10.1145/2908080.2908117}
\showDOI{\tempurl}


\bibitem[\protect\citeauthoryear{Lam, Sethi, Ullman, and Aho}{Lam
  et~al\mbox{.}}{2007}]%
        {Lam2007}
\bibfield{author}{\bibinfo{person}{M Lam}, \bibinfo{person}{R Sethi},
  \bibinfo{person}{JD Ullman}, {and} \bibinfo{person}{A Aho}.}
  \bibinfo{year}{2007}\natexlab{}.
\newblock \bibinfo{booktitle}{\emph{{Compilers: Principles, Techniques, and
  Tools}}}.
\newblock 1038 pages.
\newblock
\showISBNx{0321486811}


\bibitem[\protect\citeauthoryear{Lattner and Adve}{Lattner and Adve}{2004}]%
        {Lattner}
\bibfield{author}{\bibinfo{person}{Chris Lattner} {and} \bibinfo{person}{Vikram
  Adve}.} \bibinfo{year}{2004}\natexlab{}.
\newblock \showarticletitle{{LLVM: A compilation framework for lifelong program
  analysis \& transformation}}. In \bibinfo{booktitle}{\emph{International
  Symposium on Code Generation and Optimization (CGO)}}.
  \bibinfo{pages}{75--86}.
\newblock


\bibitem[\protect\citeauthoryear{Lattner, Amini, Bondhugula, Cohen, Davis,
  Pienaar, Riddle, Shpeisman, Vasilache, and Zinenko}{Lattner
  et~al\mbox{.}}{2021}]%
        {Lattner2021}
\bibfield{author}{\bibinfo{person}{Chris Lattner}, \bibinfo{person}{Mehdi
  Amini}, \bibinfo{person}{Uday Bondhugula}, \bibinfo{person}{Albert Cohen},
  \bibinfo{person}{Andy Davis}, \bibinfo{person}{Jacques Pienaar},
  \bibinfo{person}{River Riddle}, \bibinfo{person}{Tatiana Shpeisman},
  \bibinfo{person}{Nicolas Vasilache}, {and} \bibinfo{person}{Oleksandr
  Zinenko}.} \bibinfo{year}{2021}\natexlab{}.
\newblock \showarticletitle{{MLIR: Scaling Compiler Infrastructure for Domain
  Specific Computation}}. In \bibinfo{booktitle}{\emph{CGO 2021 - Proceedings
  of the 2021 IEEE/ACM International Symposium on Code Generation and
  Optimization}}. \bibinfo{pages}{2--14}.
\newblock
\showISBNx{9781728186139}
\urldef\tempurl%
\url{https://doi.org/10.1109/CGO51591.2021.9370308}
\showDOI{\tempurl}


\bibitem[\protect\citeauthoryear{Lee, Johnson, and Eigenmann}{Lee
  et~al\mbox{.}}{2004}]%
        {Lee2004}
\bibfield{author}{\bibinfo{person}{Sang~Ik Lee}, \bibinfo{person}{Troy~A.
  Johnson}, {and} \bibinfo{person}{Rudolf Eigenmann}.}
  \bibinfo{year}{2004}\natexlab{}.
\newblock \showarticletitle{{Cetus - An extensible compiler infrastructure for
  source-to-source transformation}}.
\newblock \bibinfo{journal}{\emph{Lecture Notes in Computer Science (including
  subseries Lecture Notes in Artificial Intelligence and Lecture Notes in
  Bioinformatics)}}  \bibinfo{volume}{2958} (\bibinfo{year}{2004}),
  \bibinfo{pages}{539--553}.
\newblock
\showISBNx{9783540246442}
\showISSN{16113349}
\urldef\tempurl%
\url{https://doi.org/10.1007/978-3-540-24644-2_35}
\showDOI{\tempurl}


\bibitem[\protect\citeauthoryear{{LLVM Documentation}}{{LLVM
  Documentation}}{2021}]%
        {LLVMDocumentation2021}
\bibfield{author}{\bibinfo{person}{{LLVM Documentation}}.}
  \bibinfo{year}{2021}\natexlab{}.
\newblock \bibinfo{title}{{TableGen}}.
\newblock
\newblock
\urldef\tempurl%
\url{https://llvm.org/docs/TableGen/}
\showURL{%
\tempurl}


\bibitem[\protect\citeauthoryear{L{\"{u}}cke, Steuwer, and Smith}{L{\"{u}}cke
  et~al\mbox{.}}{2021}]%
        {Lucke2021}
\bibfield{author}{\bibinfo{person}{Martin L{\"{u}}cke}, \bibinfo{person}{Michel
  Steuwer}, {and} \bibinfo{person}{Aaron Smith}.}
  \bibinfo{year}{2021}\natexlab{}.
\newblock \showarticletitle{{Integrating a functional pattern-based IR into
  MLIR}}. In \bibinfo{booktitle}{\emph{CC 2021 - Proceedings of the 30th ACM
  SIGPLAN International Conference on Compiler Construction}}.
  \bibinfo{pages}{12--22}.
\newblock
\showISBNx{9781450383257}
\urldef\tempurl%
\url{https://doi.org/10.1145/3446804.3446844}
\showDOI{\tempurl}


\bibitem[\protect\citeauthoryear{Moses, Chelini, Zhao, and Zinenko}{Moses
  et~al\mbox{.}}{2021}]%
        {Moses2021}
\bibfield{author}{\bibinfo{person}{William~S. Moses}, \bibinfo{person}{Lorenzo
  Chelini}, \bibinfo{person}{Ruizhe Zhao}, {and} \bibinfo{person}{Oleksandr
  Zinenko}.} \bibinfo{year}{2021}\natexlab{}.
\newblock \showarticletitle{Polygeist: Raising C to Polyhedral MLIR}.
\newblock \bibinfo{journal}{\emph{2021 30th International Conference on
  Parallel Architectures and Compilation Techniques (PACT)}},
  \bibinfo{pages}{45--59}.
\newblock
\showISBNx{978-1-6654-4278-7}
\urldef\tempurl%
\url{https://doi.org/10.1109/PACT52795.2021.00011}
\showDOI{\tempurl}


\bibitem[\protect\citeauthoryear{Narayan and Pouchet}{Narayan and
  Pouchet}{2012}]%
        {Narayan2012}
\bibfield{author}{\bibinfo{person}{Mohanish Narayan} {and}
  \bibinfo{person}{Louis-Noel Pouchet}.} \bibinfo{year}{2012}\natexlab{}.
\newblock \bibinfo{title}{{PolyBench/Fortran 1.0}}.
\newblock
\newblock
\urldef\tempurl%
\url{http://web.cse.ohio-state.edu/~pouchet.2/software/polybench/polybench-fortran.html}
\showURL{%
\tempurl}


\bibitem[\protect\citeauthoryear{Niu and Riddle}{Niu and Riddle}{2021}]%
        {PDLRegion}
\bibfield{author}{\bibinfo{person}{Jeff Niu} {and} \bibinfo{person}{River
  Riddle}.} \bibinfo{year}{2021}\natexlab{}.
\newblock \bibinfo{title}{{Public communication at LLVM/MLIR Discourse}}.
\newblock
\newblock
\urldef\tempurl%
\url{https://llvm.discourse.group/t/how-can-i-use-pdl-to-match-affine-code/4837}
\showURL{%
\tempurl}


\bibitem[\protect\citeauthoryear{{Offensive Security}}{{Offensive
  Security}}{2021}]%
        {OffensiveSecurity2021}
\bibfield{author}{\bibinfo{person}{{Offensive Security}}.}
  \bibinfo{year}{2021}\natexlab{}.
\newblock \bibinfo{title}{{The official Exploit Database repository}}.
\newblock
\newblock
\urldef\tempurl%
\url{https://github.com/offensive-security/exploitdb}
\showURL{%
\tempurl}


\bibitem[\protect\citeauthoryear{Pelegri-Llopart and Graham}{Pelegri-Llopart
  and Graham}{1988}]%
        {Pelegri-Llopart1988}
\bibfield{author}{\bibinfo{person}{Eduardo Pelegri-Llopart} {and}
  \bibinfo{person}{Susan~L Graham}.} \bibinfo{year}{1988}\natexlab{}.
\newblock \showarticletitle{{Optimal code generation for expression trees: An
  application of BURS theory}}. In \bibinfo{booktitle}{\emph{Conference Record
  of the Annual ACM Symposium on Principles of Programming Languages}},
  Vol.~\bibinfo{volume}{Part F1301}. \bibinfo{pages}{294--308}.
\newblock
\showISBNx{0897912527}
\showISSN{07308566}
\urldef\tempurl%
\url{https://doi.org/10.1145/73560.73586}
\showDOI{\tempurl}


\bibitem[\protect\citeauthoryear{Pottenger and Eigenmann}{Pottenger and
  Eigenmann}{1995}]%
        {Pottenger1995}
\bibfield{author}{\bibinfo{person}{Bill Pottenger} {and}
  \bibinfo{person}{Rudolf Eigenmann}.} \bibinfo{year}{1995}\natexlab{}.
\newblock \showarticletitle{{Idiom recognition in the polaris parallelizing
  compiler}}. In \bibinfo{booktitle}{\emph{Proceedings of the International
  Conference on Supercomputing}}, Vol.~\bibinfo{volume}{Part F1293}.
  \bibinfo{pages}{444--448}.
\newblock
\showISBNx{0897917286}
\urldef\tempurl%
\url{https://doi.org/10.1145/224538.224655}
\showDOI{\tempurl}


\bibitem[\protect\citeauthoryear{Redmon}{Redmon}{2016}]%
        {Redmon2016}
\bibfield{author}{\bibinfo{person}{Joseph Redmon}.}
  \bibinfo{year}{2016}\natexlab{}.
\newblock \bibinfo{title}{{Darknet: Open Source Neural Networks in C}}.
\newblock
\newblock
\urldef\tempurl%
\url{http://pjreddie.com/darknet/}
\showURL{%
\tempurl}


\bibitem[\protect\citeauthoryear{repository}{repository}{2021a}]%
        {CILrepository2021}
\bibfield{author}{\bibinfo{person}{CIL repository}.}
  \bibinfo{year}{2021}\natexlab{a}.
\newblock \bibinfo{title}{{Commit}}.
\newblock
\newblock
\urldef\tempurl%
\url{https://github.com/compiler-tree-technologies/cil/commit/195acc33e14715da9f5a1746b489814d56c015f7}
\showURL{%
\tempurl}


\bibitem[\protect\citeauthoryear{repository}{repository}{2021b}]%
        {FIRrepository2021}
\bibfield{author}{\bibinfo{person}{FIR repository}.}
  \bibinfo{year}{2021}\natexlab{b}.
\newblock \bibinfo{title}{{Commit}}.
\newblock
\newblock
\urldef\tempurl%
\url{https://github.com/flang-compiler/f18-llvm-project/commit/8abd290c2c791c26cd1237b218def1b85998d403}
\showURL{%
\tempurl}


\bibitem[\protect\citeauthoryear{repository}{repository}{2021c}]%
        {LLVM/MLIRrepository2021}
\bibfield{author}{\bibinfo{person}{LLVM/MLIR repository}.}
  \bibinfo{year}{2021}\natexlab{c}.
\newblock \bibinfo{title}{{Commit}}.
\newblock
\newblock
\urldef\tempurl%
\url{https://github.com/llvm/llvm-project/commit/1fdec59bffc11ae37eb51a1b9869f0696bfd5312}
\showURL{%
\tempurl}


\bibitem[\protect\citeauthoryear{Riddle}{Riddle}{2021}]%
        {Riddle2021}
\bibfield{author}{\bibinfo{person}{River Riddle}.}
  \bibinfo{year}{2021}\natexlab{}.
\newblock \bibinfo{title}{{Pattern Descriptor Language}}.
\newblock
\newblock
\urldef\tempurl%
\url{https://drive.google.com/file/d/17WYUvlmCzNTiqLaxWf_uz4GiLm3QVoEV/view}
\showURL{%
\tempurl}


\bibitem[\protect\citeauthoryear{Xianyi and Kroeker}{Xianyi and
  Kroeker}{2020}]%
        {Xianyi2020}
\bibfield{author}{\bibinfo{person}{Zhang Xianyi} {and} \bibinfo{person}{Martin
  Kroeker}.} \bibinfo{year}{2020}\natexlab{}.
\newblock \bibinfo{title}{{OpenBLAS: An optimized BLAS library}}.
\newblock
\newblock
\urldef\tempurl%
\url{https://www.openblas.net/}
\showURL{%
\tempurl}


\bibitem[\protect\citeauthoryear{Zemlyachenko, Korneenko, and
  Tyshkevich}{Zemlyachenko et~al\mbox{.}}{1985}]%
        {dagi}
\bibfield{author}{\bibinfo{person}{V~N Zemlyachenko}, \bibinfo{person}{N~M
  Korneenko}, {and} \bibinfo{person}{R~I Tyshkevich}.}
  \bibinfo{year}{1985}\natexlab{}.
\newblock \showarticletitle{{Graph isomorphism problem}}.
\newblock \bibinfo{journal}{\emph{Journal of Soviet Mathematics}}
  \bibinfo{volume}{29}, \bibinfo{number}{4} (\bibinfo{year}{1985}),
  \bibinfo{pages}{1426--1481}.
\newblock
\showISSN{1573-8795}
\urldef\tempurl%
\url{https://doi.org/10.1007/BF02104746}
\showDOI{\tempurl}


\end{thebibliography}

\clearpage

%
%
\setcounter{section}{0}

\section{Artifact Appendix}
\label{sec:appendix}

\subsection{Abstract}

SMR's artifact reproduces the results presented in \rsec{results}, thus validating  the algorithms' aspects discussed throughout the paper. It is composed of a Docker container with bash scripts that reproduce the paper's results (\rfig{res_poly_Fortran} and \rfig{compilation_time}). A Linux system with an up-to-date Docker installation and at least 20G of disk memory is required. One of the results (\rtab{c-matches}) is not entirely reproducible due to its computational heft, but can be partially executed.

\subsection{Artifact check-list (meta-information)}

\begin{itemize}
    \item \textbf{Algorithm:} Source Matching and Rewriting (SMR), a source code pattern matching and rewriting algorithm based on the MLIR framework.
    \item \textbf{Program:} Kernels utilized in the artifact are from PolyBench/Fortran 1.0 benchmark (already included).
    \item \textbf{Data set:} \texttt{LARGE\_DATASET} mode predefined in PolyBench/Fortran 1.0.
    \item \textbf{Run-time environment:} Linux system supported by LLVM with up-to-date Docker support (Ubuntu 18 or greater is recommended).
    \item \textbf{Output:} PDF files replicating \rfig{res_poly_Fortran} and \rfig{compilation_time}, which show SMR rewrite optimization speedups and compilation overhead.
    \item \textbf{Disk space required:} $\approx 20$ gigabytes of free space.
    \item \textbf{Time to run experiments:} $\approx 20$ minutes.
    \item \textbf{Publicly available:} Yes, via GitLab and Docker hub.
\end{itemize}

\subsection{Description}

\subsubsection{How is it delivered?}
\begin{itemize}
    \item A container with all required tools can be downloaded with \texttt{docker pull sitio/smr-artifact}
    \item A Dockerfile and all other requirements to build the container can be found at the artifact's repository~\cite{smr-artifact-repo}.
\end{itemize}

\subsubsection{Hardware dependencies:}
\begin{itemize}
    \item Any platform supported by LLVM~\cite{llvm-hardware}
\end{itemize}

\subsubsection{Software dependencies:}

All required binaries are contained within the docker container.

\begin{itemize}
    \item clang, cmake, flang, gfortran and libopenblas-dev can be installed via Ubuntu's package manager.
    \item FIR is available at f18-llvm-project/fir-dev@8abd29~\cite{fir}.
    \item MLIR 11 is available at llvm-project/tree/release/11.x~\cite{LLVM/MLIRrepository2021}.
    \item PolyBench/Fortran is available at www.cs.colostate.edu~\cite{Narayan2012}.
    \item CIL is available at compiler-tree-technologies/cil~\cite{cil}
    \item SMR is available at parlab/pat-compiler~\cite{smr-binary-repo}.
\end{itemize}

\subsection{Experiment workflow}

Bash scripts can be used to reproduce the experiments:

\begin{itemize}
    \item \texttt{execution\_times.sh} to reproduce \rfig{res_poly_Fortran}
    \item \texttt{compilation\_times.sh} to reproduce \rfig{compilation_time}
    \item \texttt{angha\_matches.sh} to \textbf{partially} reproduce \rtab{c-matches}
\end{itemize}

\par To demonstrate correctness, an extra script (\texttt{validate.sh}) can be used to compare PolyBench reference values against the rewrites applied by SMR in the \texttt{execution\_times.sh} experiment.

\par Steps to reproduce the experiments:

\begin{itemize}
    \item \texttt{docker pull sitio/smr-artifact}
    \item \texttt{docker run -it --name smr sitio/smr-artifact}
    \item \texttt{cd /root/smr-artifact}
    \item \texttt{bash compilation\_times.sh}
    \item \texttt{bash execution\_times.sh}
    \item \texttt{bash angha\_matches.sh}
    \item \texttt{bash validate.sh}
\end{itemize}

After executing the steps above, two PDF files will be generated in \texttt{/root/smr-artifact}: \texttt{compilation\_times.pdf} and \texttt{execution\_times.pdf}. In order to visualize the graphs, copy the generated PDFs files from the container to the host machine with \href{https://docs.docker.com/engine/reference/commandline/cp}{\texttt{docker cp}} and open them with any PDF viewer.

Results for the \texttt{angha\_matches.sh} and \texttt{validate.sh} scripts are exhibited directly on \texttt{stdout}. The first will list how many matches were found for each C file in the \texttt{angha} folder, and the latter will log if the SMRs results are within and acceptable relative error margin.

\subsection{Evaluation and expected results}

\par Experiment \texttt{compilation\_times.sh} shows the overhead added by the entire SMR rewriting process. The column representing SMR is expected to be taller (slower) in every kernel, how much taller will depend of the host machine.

\par Experiment \texttt{execution\_times.sh} quantifies the improvement achieved by rewriting the traditional PolyBench kernels by its BLAS counterparts using SMR. This experiment can also vary depending on the host machine, however, times for gFortran and Flang binaries should always be substantially larger when compared with SMR and BLAS times.

\par Experiment \texttt{angha\_matches.sh} shows SMR functioning with the CIL dialect and frontend, reiterating its flexibility for multiple source languages. However, this experiment is only partially reproduced in the artifact: the web crawling, code reconstruction and input code filtering steps are all left out due to their computational demanding nature, which would take several hours to reproduce.

\par The \texttt{validation.sh} script checks correctness of the rewrites performed by SMR with BLAS in the \texttt{execution\_times.sh} experiment. The validation consists of a relative error check which tolerates a margin of 0.00001\% from PolyBench reference values. Any result that exceeds the threshold will be printed as an error in \texttt{stdout}.

\end{document}